\documentclass{manuscript}
\title{SemanticCAP: Chromatin Accessibility Prediction Enhanced by Features Learning from a Language Model}
\author{Yikang Zhang, Xiaomin Chu, Yelu Jiang, Hongjie Wu and Lijun Quan}

\begin{document}
	\mabstract{A large number of inorganic and organic compounds are able to bind DNA and form complexes, among which drug-related molecules are important. Chromatin accessibility changes not only directly affects drug-DNA interactions, but also promote or inhibit the expression of critical genes associated with drug resistance by affecting the DNA binding capacity of TFs and transcriptional regulators. However, Biological experimental techniques for measuring it are expensive and time consuming. In recent years, several kinds of computational methods have been proposed to identify accessible regions of the genome. Existing computational models mostly ignore the contextual information of bases in gene sequences. To address these issues, we proposed a new solution named SemanticCAP. It introduces a gene language model which models the context of gene sequences, thus being able to provide an effective representation of a certain site in gene sequences. Basically, we merge the features provided by the gene language model into our chromatin accessibility model. During the process, we designed some methods to make feature fusion smoother. Compared with other systems under public benchmarks, our model proved to have better performance.}{chromatin accessibility; drug design; language model; transformer; feature fusion}
	
	\section{Introduction}
In human cells, genetic and regulatory information is stored in chromatin, which is deoxyribonucleic acid (DNA) wrapped around histones. Chromatin structure has a lot to do with gene transcription, protein synthesis, biochemical processes and other complex biological expressions. Among them, binding of small organic and inorganic molecules to DNA can influence numerous biological processes in which DNA participate. In particular, many anticancer, antibiotic, and antiviral drugs exert their primary biological effects by reversibly interacting with nucleic acids. Therefore, the study of its structure can help us design drugs to control gene expression and cure diseases \citep{aleksic2014overview}. Some regions of the chromatin are open to transcription factors (TFs), RNA polymers (RNAPs), drug molecules and other cellular materials, while others are tightly entangled together and do not play a part in most cellular processes. These two regions of the chromatin are called open regions and closed regions, which are also known as accessibility and inaccessibility separately \citep{wang2016modeling}. Measuring it can generate clues to gene function that help to identify appropriate targets for therapeutic intervention. Meanwhile, monitoring changes in chromatin accessibility can help track and understand drug effects. \citet{gallon2021chromatin} find that chromatin accessibility changes at intergenic regions are associated with ovarian cancer drug resistance. Another example is the chromatin opening (increased accessibility) of the targeted DNA satellites can explain how DNA-binding pyrrole-imidazole compounds that target different \textit{Drosophila melanogaster} satellites leading to gain- or loss-of-function phenotypes \citep{janssen2000specific}. In recent years, many high-throughput sequencing technologies have been used for the detection of open regions, such as DNase-seq \citep{song2010dnase}, FAIRE-seq \citep{simon2012using} and ATAC-seq \citep{buenrostro2015atac}. However, biological experimental methods are costly and time-consuming, thus cannot be applied to large-scale chemical examinations. These restrictions have promoted the development of calculation methods.
\par 
Alongside the progress in computer science, several kinds of sequence-based calculation methods have been proposed to identify functional regions. Simply we can divide them into traditional machine learning methods \citep{lee2011discriminative,ghandi2014enhanced,beer2017predicting,xu2021integration,kumar2016predicting} and neural network methods \citep{alipanahi2015predicting,zhou2015predicting,min2017chromatin,liu2018chromatin,guo2020deepanf}. Machine learning methods are mainly based on support vector machines (SVM). \citet{lee2011discriminative} designed a SVM method based on k-mer features, which is defined as a segment of length $k$. This method recognizes enhancers in mammalian cells. Subsequently, the gkm-SVM (gapped k-mers SVM) posed by \citet{ghandi2014enhanced} exploited a feature set called interval k-mer features to improve the accuracy and stability of recognition. In recent years, with the rapid development of neural networks and the emergence of various deep learning models, a growing number of deep network models are used to solve such problems, where convolutional neural networks (CNNs) \citep{kalchbrenner2014convolutional} and recurrent neural networks (RNNs) \citep{hochreiter1997long} dominate. DeepBind \citep{alipanahi2015predicting} in 2015 and DeepSEA \citep{zhou2015predicting} in 2017 got a significant performance improvement compared with traditional SVM-based methods after applying CNNs to modeling the sequence specificity of protein binding. \citet{min2017chromatin} utilized the long short-term memory network (LSTM) to predict the chromatin accessibility and achieved state-of-the-art at the time, thus proving the effectiveness of RNNs for DNA sequence problems.
\par
However, we point out that the previous methods have the following shortcomings at least. First, most of the previous methods are based on k-mer, that is, a segment of length $k$. Specifically, it takes a segment of length $k$ at intervals. The artificial division of the original sequence may destroy the inside semantic information and cause difficulty in the learning of subsequent models. Second, with the progress of language models, we have the ability to learn the interior semantic information of sequences through pre-training. There have been related works in existed methods, such as \citet{min2017chromatin} using Glove \citep{pennington2014glove} to train the word vectors of k-mer. However, these pre-train models are mostly traditional word vector methods. On the one hand, they can only learn the characteristics of the word itself and have no knowledge of the context of DNA sequences \citep{sun2019deep}. On the other hand, they are limited to a specific dataset thus cannot be widely applied to other scenarios. Third, traditional CNNs and RNNs have been proved to be unsuitable for long-sequence problems \citep{bengio2013representation}. CNNs restricted by the size of convolution kernels fail in learning global information effectively, while RNNs tend to cause gradient disappearance and slow training due to the lack of parallelizability when getting a long input. In contrast, the attention mechanism (Attention) \citep{vaswani2017attention} can effectively learn the long-range dependence of sequences, which has been widely used in the field of natural language processing.
\par
In response to the above disadvantages, we constructed a chromatin accessibility prediction model based on features learning from a language model. A implemention of our system is available at \href{https://www.github.com/ykzhang0126/SemanticCAP}{github.com/ykzhang0126/SemanticCAP}. Our method has at least the following three improvements:
\begin{enumerate}
	\item A DNA language model is utilized to learn the deep semantics of DNA sequences, and introduces the semantic features in the process of chromatin accessibility prediction. Therefore we obtain additional complex environmental information.
	\item Both the DNA language model and the chromatin accessibility model use character-based input instead of k-mer which stand for segments of length $k$. The strategy prevents the information of original sequences from being destroyed.
	\item The attention mechanism is widely used in our models in place of CNNs and RNNs, which can make the model more powerful and stable in handling long sequences.
\end{enumerate}
\par
Before formally introducing our method, we first present some preliminary knowledge, including some common sense information, theorems and corollaries.
	\section{Theories}

\begin{Theorem}
	\label{theorem1}
	For two standardized distributions using layer normalization (LN), which are denoted as $X_1$ and $X_2$, the concat of them, that is, $X\equiv[X_1,X_2]$, is still a standardized distribution.
\end{Theorem}
\begin{proof}
	Suppose that $X_1$ has $n$ elements and $X_2$ has $m$ elements. As we all know, LN \citep{ba2016layer} transforms the distribution $X$ as
	\begin{equation}
		\label{c2.t1.eq1}
		X\xrightarrow[]{\mathrm{LN}} \frac{X-\mu}{\sigma}
	\end{equation}
	where $\mu$ and $\sigma$ are the expectation and standard deviation of $X$, respectively. Obviously, for the normalized distribution $X_1$ and $X_2$, we have
	\begin{equation}
		\label{c2.t1.eq2}
		E(X_1)=E(X_2)=0
	\end{equation}
	\begin{equation}
		\label{c2.t1.eq3}
		D(X_1)=D(X_2)=1
	\end{equation}
	where $E$ stands for the expectation function and $D$ stands for the deviation function. The new distribution $X$ is derived by concating $X_1$ and $X_2$, thus having $n+m$ elements. Inferring from Eq. \ref{c2.t1.eq2}, we get
	\begin{equation}
		\label{c2.t1.eq4}
		E(X)=\frac{nE(X_1)+mE(X_2)}{n+m}=0
	\end{equation}
	\begin{equation}
		\label{c2.t1.eq5}
		\begin{aligned}
			D(X)&=E(X^2)-E^2(X)=E(X^2)\\
			&=\frac{nE(X_1^2)+mE(X_2^2)}{n+m}
		\end{aligned}
	\end{equation}
	For $X_1$ and $X_2$, we also know that
	\begin{equation}
		\label{c2.t1.eq6}
		E(X_1^2)=D(X_1)+E^2(X_1)
	\end{equation}
	\begin{equation}
		\label{c2.t1.eq7}
		E(X_2^2)=D(X_2)+E^2(X_2)
	\end{equation}
	Substituting Eq. \ref{c2.t1.eq2} and Eq. \ref{c2.t1.eq3} into Eq. \ref{c2.t1.eq6} and Eq. \ref{c2.t1.eq7}, and finally into Eq. \ref{c2.t1.eq5}, we get
	\begin{equation}
		\label{c2.t1.eq8}
		D(X)=1
	\end{equation}
	Eq. \ref{c2.t1.eq4} and Eq. \ref{c2.t1.eq8} demonstrate the standardization of $X$.
\end{proof}

\begin{Theorem}
	\label{theorem2}
	For any two distributions $X_1$, $X_2$, there always exist two coefficients $\lambda_1$, $\lambda_2$, so that the concat of them after being multiplied by the two coefficients respectively, that is $X\equiv[\lambda_1X_1,\lambda_2X_2]$, is a standardized distribution.
\end{Theorem}
\begin{proof}
	Suppose that $X_1$ has $n$ elements and $X_2$ has $m$ elements. We denote the expectation of the two distributions as $\mu_1$, $\mu_2$, and the variance as $\sigma_1^2$, $\sigma_2^2$. Notice that $\lambda_1$ and $\lambda_2$ are all scalars. Now pay attention to $X$. To prove this theorem, we want $X$ to be a standardized distribution, which requires the expectation of $X$ to be $0$ and the variance to be $1$. Therefore, we can list the following equation set.
	\begin{equation}
		\label{c2.t2.eq1}
		\left\{
			\begin{aligned}
				&E(X)=\frac{n\lambda_1E(X_1)+m\lambda_2E(X_2)}{n+m}=0\\
				&E(X^2)=\frac{nE((\lambda_1X_1)^2)+mE((\lambda_2X_2)^2)}{n+m}\\
				&D(X)=E(X^2)-E^2(X)=1
			\end{aligned}
		\right.
	\end{equation}
	At the same time, we have equations similar to Eq. \ref{c2.t1.eq6} and Eq. \ref{c2.t1.eq7}, that is
	\begin{equation}
		\label{c2.t2.eq2}
			E((\lambda_1X_1)^2)=D(\lambda_1X_1)+E^2(\lambda_1X_1)
	\end{equation}
	\begin{equation}
		\label{c2.t2.eq3}
		E((\lambda_2X_2)^2)=D(\lambda_2X_2)+E^2(\lambda_2X_2)
	\end{equation}
	which is easy to calculate according to the nature of expectation and variance. Notice that Eq. \ref{c2.t2.eq1} has two variables and two independent equations, meaning it should be solvable. By calculating Eq. \ref{c2.t2.eq1}, we can get the numeric solution of $\lambda_1$ and $\lambda_2$ as follows.
	\begin{equation}
		\label{c2.t2.eq4}
		\left\{
			\begin{aligned}
				&\lambda_1^2=\frac{m(n+m)\mu_2^2}{nm\mu_2^2(\mu_1^2+\sigma_1^2)+n^2\mu_1^2(\mu_2^2+\sigma_2^2)}\\
				&\lambda_2^2=\frac{n(n+m)\mu_1^2}{nm\mu_1^2(\mu_2^2+\sigma_2^2)+m^2\mu_2^2(\mu_1^2+\sigma_1^2)}
			\end{aligned}
		\right.
	\end{equation}
	The existence of Eq. \ref{c2.t2.eq4} ends our proof. Actually, we get two sets of solutions here, for $\lambda_1$ can be either positive or negative, and so be $\lambda_2$. The signs of $\lambda_1$ and $\lambda_2$ depend on the signs of $\mu_1$ and $\mu_2$, which can be easily inferred from the first equation in Eq. \ref{c2.t2.eq1}.
\end{proof}
\begin{Corollary}
	For any distributions $X_1$, $X_2$, $\dots$, $X_n$, there always exist $n$ coefficients $\lambda_1$, $\lambda_2$, $\dots$, $\lambda_n$, so that the concat of them after being multiplied by the $n$ coefficients respectively, that is $X\equiv[\lambda_1X_1,\lambda_2X_2,\dots,\lambda_nX_n]$, is a standardized distribution.
\end{Corollary}
\begin{proof}
	The overall proof way is similar to that in Theorem \ref{theorem2}. Note that in this case, we have $n$ variables but only $2$ independent equations, resulting in infinite solutions according to Eq. \ref{c2.t2.eq1}. To be more precise, the degree of freedom of our solutions is $n-2$.
\end{proof}

\begin{Theorem}
	\label{theorem3}
	In neural networks, for any two tensors $X$, $Y$ which satisfy $E(X)=E(Y)=0$, the probability of feature disappearance of $X$ after concating and normalizing them is $\Omega(\frac{SD(Y)}{SD(X)})$, where $SD$ represents the standard deviation.
\end{Theorem}
\begin{proof}
	Feature disappearance is defined as a situation where the features are too small. Concretely, for a tensor $X$ and a threshold $t_{FD}$, if the result of a subsequent operation of $X$ is smaller than $t_{FD}$, the feature disappearance of $X$ happens. Here $t_{FD}$ can be an arbitrarily small value, such as $10^{-5}$.
	\par
	Suppose that $X$ has $n$ elements and $Y$ has $m$ elements. We denote the expectation of the two distributions as $\mu_1$, $\mu_2$, and the variance as $\sigma_1^2$, $\sigma_2^2$. As stated in the precondition, we already know
	\begin{equation}
		\label{c2.t3.eq1}
		\mu_1=\mu_2=0
	\end{equation}
	Let $Z\equiv[X,Y]$ and $Z'\equiv \mathrm{LN}(Z)\equiv[X',Y']$. By the help of Eq. \ref{c2.t2.eq1}, we have
	\begin{equation}
		\label{c2.t3.eq2}
		E(Z)=0
	\end{equation}
	\begin{equation}
		\label{c2.t3.eq3}
		D(Z)=E(Z^2)=\frac{n\sigma_1^2+m\sigma_2^2}{n+m}
	\end{equation}
	We denote $E(Z)$ as $\mu$ and $D(Z)$ as $\sigma^2$. According to Eq. \ref{c2.t1.eq1}, for $X'$ we know that
	\begin{equation}
		\label{c2.t3.eq4}
		E(X')=E(\frac{X-E(Z)}{\sqrt{D(Z)}})=\frac{\mu_1-\mu}{\sigma}
	\end{equation}
	\begin{equation}
		\label{c2.t3.eq5}
		D(X')=D(\frac{X-E(Z)}{\sqrt{D(Z)}})=\frac{\sigma_1^2}{\sigma^2}
	\end{equation}
	We denote $E(X')$ as $\mu_1'$ and $D(X')$ as $\sigma_1'^2$. Now we consider the result of a subsequent operation of $X$, which is $\sum_{i=1}^{n} \lambda_iX_i'$. This is very common in convolution, linear or attention layers. For the result, an observation is
	\begin{equation}
		\label{c2.t3.eq6}
		\begin{aligned}
			\sum_{i=1}^{n} \lambda_iX_i'\le\sum_{i=1}^{n} |\lambda_i||X_i'|\le\lambda_{m}\sum_{i=1}^{n}|X_i'|
		\end{aligned}
	\end{equation}
	where $\lambda_m=\mathop{\max}\limits_{1\le i \le n}|\lambda_i|$. For the convenience of analysis, all $\lambda$ are set to 1. This will not bring loss of generality, because the value scaling from $\lambda_m$ to 1 has no effect in the subsequent derivation. Here we denote $\sum X'$ as $S_{X'}$. According to the central limit theorem (Lindeberg-L\'evy form) \citep{gine1983levy}, we find that $S_{X'}$ obeys a normal distribution, that is
	\begin{equation}
		\label{c2.t3.eq7}
		S_{X'}\sim \mathcal{N}(n\mu_1',n\sigma_1'^2)
	\end{equation}
	For a feature disappearance threshold $t_{FD}$, We want to figure out the probability of $|S_{X'}|< t_{FD}$. Denote this event as $FD$ and we can get
	\begin{equation}
		\label{c2.t3.eq8}
		\begin{aligned}
			P(FD)&=Pr\{|S_{X'}|<t_{FD}\}\\
			&=Pr\{|\frac{S_{X'}-n\mu_1'}{\sqrt{n}\sigma_1'}|<\frac{t_{FD}-n\mu_1'}{\sqrt{n}\sigma_1'}\}\\
			&=2\Phi(\frac{t_{FD}-n\mu_1'}{\sqrt{n}\sigma_1'})-1
		\end{aligned}
	\end{equation}
	where $\Phi$ is the cumulative distribution function (cdf) of the standard normal distribution. Since it is an integral which does not have a closed form solution, we cannot directly analyze it. According to Eq. \ref{c2.t3.eq1}, Eq. \ref{c2.t3.eq2} and Eq. \ref{c2.t3.eq4}, we know $\mu_1'=0$. At the same time we know that $t_{FD}$ is a small number, leading to $\frac{t_{FD}}{\sqrt{n}\sigma_1'}\to0$. Therefore, we have the equation as follows.
	\begin{equation}
		\label{c2.t3.eq9}
		\begin{aligned}
			\Phi(x)&=\Phi(0)+\varphi(0)x+R_1(x)\\
			&=0.5+\frac{1}{\sqrt{2\pi}}x+o(x)
		\end{aligned}
	\end{equation}
	The formula is a Taylor expansion where $\varphi$ is the probability density function (pdf) of the standard normal distribution, $R_1(x)$ is the Lagrange remainder and $o(x)$ is the Peano remainder standing for a high-order infinitesimal of $x$. Combining Eq. \ref{c2.t3.eq3}, Eq. \ref{c2.t3.eq5}, Eq. \ref{c2.t3.eq8} and Eq. \ref{c2.t3.eq9}, we cat get
	\begin{equation}
		\label{c2.t3.eq10}
		\begin{aligned}
			P(FD)&=t_{FD}\sqrt{\frac{2(k_sk_d^2+1)}{\pi n(k_s+1)}}+o(k_d)\\
			&=\Omega(K_d)
		\end{aligned}
	\end{equation}
	where $k_s=\frac{m}{n}$ and $k_d=\frac{\sigma_2}{\sigma_1}$. The above equation can also be written as $P(FD)=\Omega(\frac{SD(Y)}{SD(X)})$.
\end{proof}
\begin{Corollary}
	\label{corollary1}
	In neural networks, feature disappearance can lead to gradient disappearance.
\end{Corollary}
\begin{proof}
	According to Theorem \ref{theorem3}, the feature disappearance happens if there exists a tensor $T$ such that $|T|<t_{FD}$. Similar to the definition of feature disappearance, gradient disappearance is defined as a situation where the gradients are too small. Concretely, for a parameter $C$ with the gradient of $grad_C$ and a threshold $t_{GD}$, if $grad_C$ is smaller than $t_{GD}$, the gradient disappearance of $C$ happens. Here $t_{GD}$ can be an arbitrarily small value.
	\par
	Consider a subsequent operation of $T$, which is $T'=CT^n$, where $n$ stands for the number of layers involved in the calculation. The gradient disappearance happens if
	\begin{equation}
		\label{t2.c1.eq1}
		|grad_C|=|\frac{\dif T'}{\dif C}|=|T^n|=|T|^n<t_{GD}
	\end{equation}
	At the same time, we already have $|T|^n<t_{FD}^n$, which means we just need to meet
	\begin{equation}
		\label{t2.c1.eq2}
		t_{FD}^n<t_{GD}
	\end{equation}
	Note that $t_{FD}$ is a small number, which means $t_{FD}<1$. Finally we derive a formula of $n$.
	\begin{equation}
		\label{t2.c1.eq3}
		n>\frac{\log t_{GD}}{\log t_{FD}}
	\end{equation}
	Thereby we get a sufficient condition for $n$ and we can come to a conclusion. The disappearance of gradients occurs in layers deep enough after the disappearance of features.
\end{proof}
\par
The above corollary is consistent with intuition. The disappearance of gradients is always accompanied by the disappearance of features, and it is always a problem in deep neural networks.

\begin{Theorem}
	\label{theorem4}
	In neural networks, for any two tensors $X_1$, $X_2$ of the same dimension, there always exist two matrices $M_1$, $M_2$, so that the operation of concating them and the operation of adding them after being multiplied in the Hadamard format by the two matrices respectively, are equivalent in effect.
\end{Theorem}
\begin{proof}
	First of all, we illustrate the definition of Hadamard product \citep{horn1990hadamard}. The Hadamard product (also known as the element-wise product) is a binary operation that takes two matrices of the same dimensions and produces another matrix of the same dimension. Concretely, we can define it as
	\begin{equation}
		\label{c2.t4.eq1}
		\begin{aligned}
			\mathrm{A}(R,N)\circ\mathrm{B}&(R,N)=\mathrm{AB}(R,N)\\
			\mathrm{AB}&_{ij}=\mathrm{A}_{ij}\mathrm{B}_{ij}
		\end{aligned}
	\end{equation}
	The symbol '$\circ$' is used to distinguish from the more common matrix product, which is denoted as '$\cdot$' and usually be omitted. The definition implies that the dimension of $X_1$ should be the same as that of $M_1$, so are $X_2$ and $M_2$. At the same time $X_1$ and $X_2$ are assumed to have the same dimension in the precondition of our proposition. So we might as well set them to $\mathbb{R}^{R\times N}$. Now we give the representation of $X_1$ and $X_2$.
	\begin{equation}
		\label{c2.t4.eq2}
		X_1=\left(
		\begin{matrix}
			p_{1}\\
			p_{2}\\
			\vdots\\
			p_{R}
		\end{matrix}
		\right)
		\quad \quad
		X_2=\left(
		\begin{matrix}
			q_{1}\\
			q_{2}\\
			\vdots\\
			q_{R}
		\end{matrix}
		\right)
	\end{equation}
	\par
	Our goal is to weigh the effect of the two operations. For the convenience of comparison, we let the results after the two operations multiply a matrix, thus converting the dimension to $\mathbb{R}^{R\times M}$. Adding a linear layer is very common in neural networks, and it hardly affects the network's expression ability.
	\par
	Considering the first scheme, the concat of $X_1$ and $X_2$, we have
	\begin{equation}
		\label{c2.t4.eq3}
		\begin{aligned}
			U&=[X_1,X_2]\cdot A\\
			&=\left(
			\begin{matrix}
				p_1 & q_1\\
				p_2 & q_2\\
				\vdots & \vdots\\
				p_R & q_R
			\end{matrix}
			\right)
			\left(
			\begin{matrix}
				a_1 & a_2 & \ldots & a_M
			\end{matrix}
			\right)\\
		\end{aligned}
	\end{equation}
	where $A\in \mathbb{R}^{2N\times M}$ and $U\in \mathbb{R}^{R\times M}$. Observing the $i$-th row and $j$-th column of $U$, we can find that
	\begin{equation}
		\label{c2.t4.eq4}
		\begin{aligned}
			U_{ij}&=\left[p_i,q_i\right]\cdot a_j\\
			&=\left[p_i,q_i\right]\cdot [a_{jp},a_{jq}]\\
			&=p_i\cdot a_{jp}+q_i\cdot a_{jq}
		\end{aligned}
	\end{equation}
	\par
	Considering the second scheme, with the addition of $X_1$ and $X_2$ as the core, we have
	\begin{equation}
		\label{c2.t4.eq5}
		\begin{aligned}
			V&=(M_1\circ X_1+M_2\circ X_2)\cdot B\\
			&=\left(
			\begin{matrix}
				\lambda_{11}\circ p_1+\lambda_{12}\circ q_1\\
				\lambda_{21}\circ p_2+\lambda_{22}\circ q_2\\
				\vdots\\
				\lambda_{R1}\circ p_R+\lambda_{R2}\circ q_R
			\end{matrix}
			\right)
			\left(
			\begin{matrix}
				b_1 & b_2 & \ldots & b_M
			\end{matrix}
			\right)
		\end{aligned}
	\end{equation}
	where $B\in \mathbb{R}^{N\times M}$ and $V\in \mathbb{R}^{R\times M}$. Still we pay attention to the $i$-th row and $j$-th column of $V$ and find that
	\begin{equation}
		\label{c2.t4.eq6}
		\begin{aligned}
			V_{ij}&=(\lambda_{i1}\circ p_i+\lambda_{i2}\circ q_i)\cdot b_j\\
			&=\sum ((\lambda_{i1}\circ p_i+\lambda_{i2}\circ q_i)\circ b_j^\mathrm{T})\\
			&=\sum (\lambda_{i1}\circ b_j^\mathrm{T}\circ p_i+\lambda_{i2}\circ b_j^\mathrm{T}\circ q_i)\\
			&=\sum (\lambda_{i1}\circ b_j^\mathrm{T}\circ p_i)+\sum(\lambda_{i2}\circ b_j^\mathrm{T}\circ q_i)\\
			&=p_i\cdot(\lambda_{i1}\circ b_j^\mathrm{T})^\mathrm{T}+q_i\cdot(\lambda_{i2}\circ b_j^\mathrm{T})^\mathrm{T}\\
			&=p_i\cdot(\lambda_{i1}^\mathrm{T}\circ b_j)+q_i\cdot(\lambda_{i2}^\mathrm{T}\circ b_j)\\
		\end{aligned}
	\end{equation}
	Comparing Eq. \ref{c2.t4.eq4} and Eq. \ref{c2.t4.eq6}, we find that when we let $\lambda_{i1}^T\circ b_j$ equal to $a_{jp}$ and $\lambda_{i2}^T\circ b_j$ equal to $a_{jq}$, the values of $U$ and $V$ are equal, which is a strong evidence of effect equivalence.
\end{proof}
\par
As the equivalence has been proved, the same as the plain concat no information is lost in the above method. We point out that the Hadamard product is an alternative version of the gate mechanism \citep{liu2017gated}. We use coefficients to adjust the original distribution to screen out effective features. For the speed and stability of training, it is recommended to set the initial value of $M$ to 1. 
\par
Further, we can observe the gradient of the parameters $\lambda$ in Eq. \ref{c2.t4.eq5}, we have
\begin{equation}
	\label{c2.t4.eq7}
	\nabla 
	\left(
	\begin{matrix}
		\lambda_{11}\circ p_1+\lambda_{12}\circ q_1\\
		\lambda_{21}\circ p_2+\lambda_{22}\circ q_2\\
		\vdots\\
		\lambda_{R1}\circ p_R+\lambda_{R2}\circ q_R
	\end{matrix}
	\right)
	=
	\left(
	\begin{matrix}
		p_1 & q_1\\
		p_2 & q_2\\
		\vdots & \vdots\\
		p_R & q_R
	\end{matrix}
	\right)
\end{equation}
Compared with the gate mechanism, our method is simpler, space-saving, and more direct in gradient propagation.
\par
Of course, Theorem \ref{theorem4} could be generalized to cases with arbitrary number of tensors. We describe it in the following corollary.
\begin{Corollary}
	\label{corollary2}
	In neural networks, for any tensors $X_1$, $X_2$, $\dots$, $X_n$ of the same dimension, there always exist $n$ matrices $M_1$, $M_2$, $\dots$, $M_n$ so that the operation of concating them and the operation of adding them after being multiplied in the Hadamard format by the $n$ matrices respectively, are equivalent in effect.
\end{Corollary}
\begin{proof}
	Similar to the proof of Theorem \ref{theorem4}.
\end{proof}

\begin{Theorem}
	\label{theorem5}
	In neural networks, for a layer composed of $n$ neurons, the effective training times of the neurons in this layer reach the maximum when the dropout rate is set to $0$ or $1-\frac{1}{n}$.
\end{Theorem}
\begin{proof}
	The number of neurons in this layer is $n$, so we shall mark them as $N_1$, $N_2$, $\dots$, $N_n$. Suppose that the dropout rate \citep{baldi2013understanding} is $p$ and the total number of training times is $t$. We denote $1-p$ as q.
	\par
	Consider the $t_i$-th training. The network randomly selects $nq$ neurons for update due to the exisitence of dropout mechanism. Denote these neurons as $N_1$, $N_2$, $\dots$, $N_{nq}$.
	\par
	Without loss of generality, we consider the next time $N_1$ is selected, which is the $t_2$-th training. We denote the number of neurons selected for update in $N_2$, $\dots$, $N_{nq}$ as $S$, and the number of that selected in $N_{nq+1}$, $\dots$, $N_{n}$ as $T$. Easily we know that the selection of neurons in $S$ is an independent event, so we have
	\begin{equation}
		\label{c2.t5.eq1}
		E(S)=q(nq-1)
	\end{equation}
	At the same time, the relationship between $S$ and $T$ is
	\begin{equation}
		\label{c2.t5.eq2}
		T=nq-1-S
	\end{equation}
	Inferring from Eq. \ref{c2.t5.eq1} and Eq. \ref{c2.t5.eq2}, we get
	\begin{equation}
		\label{c2.t5.eq3}
		E(T)=-nq^2+nq+q-1
	\end{equation}
	The neurons represented by $S$ are the neurons that are updated jointly at time $t_2$ and time $t_1$, thus belong to the same sub-network. We assume that they share one training gain with $N_1$. At the same time, the neurons represented by $T$ have not been updated at time $t_1$, thus each of them has one unique training gain. Therefore, at the update time $t_2$, the expected gain of $N_1$ is $1\times \frac{1}{E(T)+1}\times \frac{1}{E(S)+1}$, which is derived from the above proportion analysis. Paying attention to $t_1$ and $t_2$, we find that $t_2-t_1$ obeys a geometric distribution, because the selection of $N_1$ is a Bernoulli experiment with probability $q$. That is $t_2-t_1\sim \mathcal{GE}(q)$, meaning that
	\begin{equation}
		\label{c2.t5.eq4}
		E(t_2-t_1)=\frac{1}{q}
	\end{equation}
	Therefore, the expected number of training times of $N_1$ is $E(\frac{t}{t_2-t_1})=tq$. The total training gain is the product of the number of training times and the gain of a single training, which we denote as $G$. Now the formula emerges.
	\begin{equation}
		\label{c2.t5.eq5}
		G(q)=\frac{t}{-n^2q^3+(n^2+2n)q^2-(2n+1)q+n+1}
	\end{equation}
	Denote $f(q)$ as the denominator of $G(q)$ and differentiate that to get
	\begin{equation}
		\label{c2.t5.eq6}
		\frac{\partial f(q)}{\partial q}=-(nq-1)(3nq-2n-1)
	\end{equation}
	With the help of Eq. \ref{c2.t5.eq6}, it is easy to draw an image of $G(p)$ shown in Figure \ref{c2.t5.fig1} where we set $t$ to $1$. The observation is that when $q=\frac{1}{n}$ or $q=1$, that is, $p$ is 0 or $1-\frac{1}{n}$, $G$ reaches the maximum value $\frac{t}{n}$, demonstrating the effective training times of $N_1$ are the largest. The conclusion can be generalized to every neuron in the layer.
	\begin{figure}[t]	
		\centering
		\includegraphics[scale=0.44]{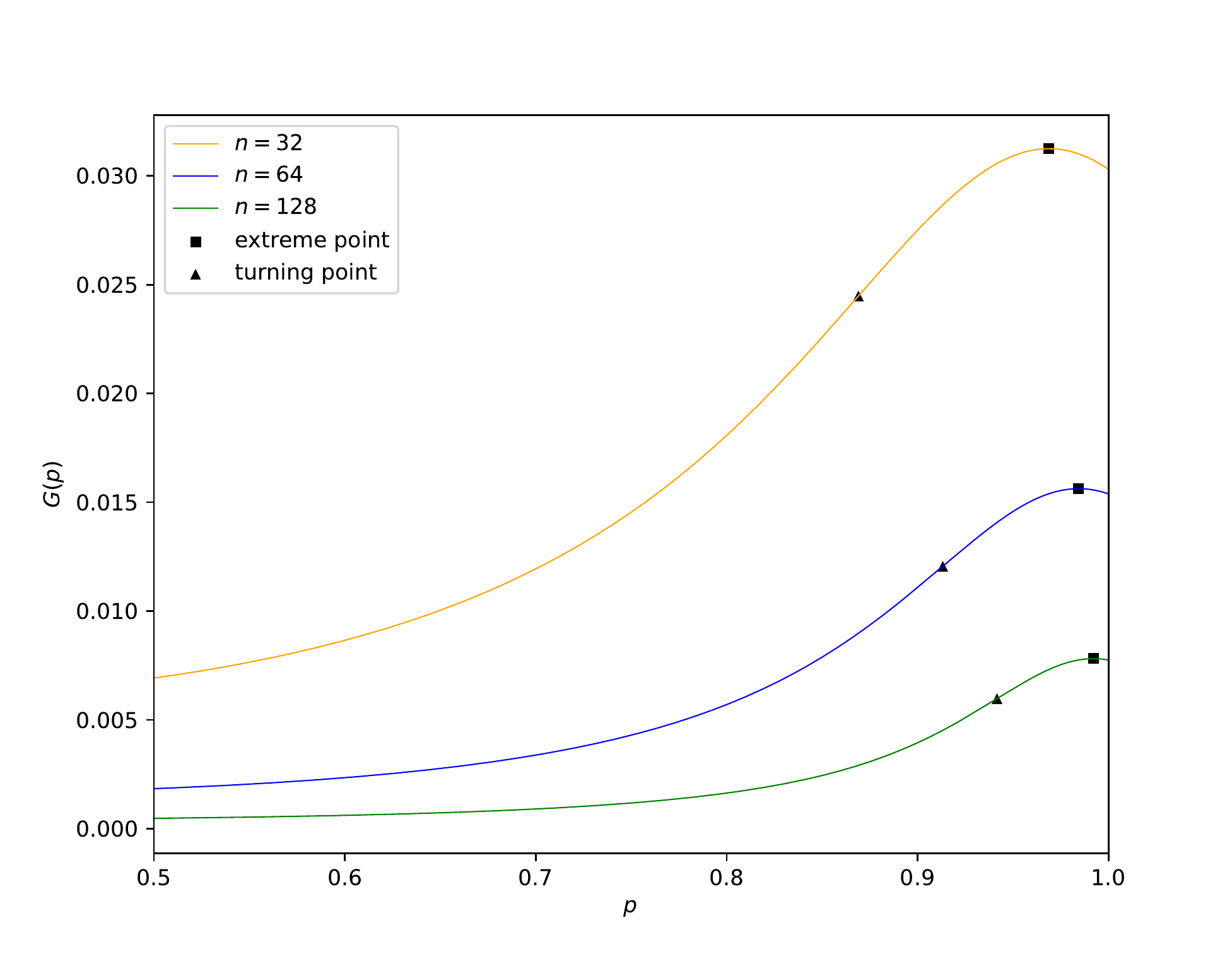}
		\caption{Gain curve as $n$ differs. $G$ varies with $p$, and the extreme points (squares) and turning points (triangles) vary with $n$. The turning point is approximately $8.72\times10^{-9}n^3 - 9.35\times10^{-6}n^2 + 2.44\times10^{-3}n + 0.78$, which is a good choice for the dropout rate.}
		\label{c2.t5.fig1}
	\end{figure}
\end{proof}
\begin{Corollary}
	\label{corollary3}
	In neural networks, if the amount of training data is sufficient, the optimal value of dropout rate is 0.5; if the amount of training data is insufficient, a number close to 1 is a better choice.
\end{Corollary}
\begin{proof}
	Theorem \ref{theorem5} focuses on the effective training times of neurons in the network, and the corollary focuses on the representation ability. It can be seen from Eq. \ref{c2.t5.eq5} that the effective training times of a certain layer is directly proportional to the total training times $t$. When the number of training times reaches a certain threshold, the network reaches a balance point, and further training will not bring any performance improvement.
	\par
	If the training data are sufficient, which means $t$ and $G$ are large enough, the network is guaranteed to get fully trained. Therefore we do not need to worry about whether the training times of neurons in the network is enough. But we still need to consider the representation ability of the network, which has a close relationship with the number of sub-networks $SN$. It can be calculated as
	\begin{equation}
		\label{c2.c2.eq1}
		SN=\left(
		\begin{matrix}
			n\\
			n(1-p)
		\end{matrix}
		\right)
	\end{equation}
	which is a combination number. Obviously, when $p$ is $0.5$, the number of sub-networks is the largest, and the network's representation ability is relatively strong.
	\par
	However, when the training data is not enough, we cannot guarantee the sufficiency of training. On the one hand, we need to set the dropout rate to a value close to $0$ or $1-\frac{1}{n}$ to guarantee the number of training as Theorem \ref{theorem5} says. On the other hand, in order to ensure the network's representation ability, we want the dropout rate be close to $0.5$. Here a balanced approach is to choose the turning point shown in Figure \ref{c2.t5.fig1}, which considers both training times and representation ability. Because this point is difficult to analyze, we give a fitting function shown in Figure \ref{c2.t5.fig1}, whose error is bounded by $2\times 10^{-2}$ for $n$ smaller than $512$.
\end{proof}
\par
The above corollary is intuitive, because the complexity of the network should be proportional to the amount of data. Little data requires a simple model, calling for a higher dropout rate. Notice that a large dropout rate not only enables the model to be fully trained, but also helps to accelerate the process. In a modern neural network framework, the discarded neurons will not participate in the gradient propagation this time, which largely reduces the number of parameters need to be adjusted in the network.

	\section{Models}
To address the task of chromatin accessibility prediction, we designed SemanticCAP, which includes a DNA language model shown in Section \ref{c3.p1} and a chromatin accessibility model shown in Section \ref{c3.p2}. Briefly, we augment the chromatin accessibility model with features provided by the DNA language model, thereby improving the performance of chromatin accessibility prediction. The detailed methods are described below.
\subsection{DNA language model}
\label{c3.p1}
\subsubsection{Process of data}
\label{c3.p1.p1}
We use the human reference genome GRCh37 (hg19) as the original data for our DNA language model. The human reference genome is a digital nucleic acid sequence database that can be used as a representative example of the gene set of an idealized individual of a species \citep{pan2019similarities}. Therefore, the model based on the database could be applied to various genetic-sequence-related issues.
\par
The task we designed for our DNA language model is using context to predict the intermediate base. However, there are at least three challenges. First is that there are two inputs, the upstream and downstream, which are defined as the upper and lower sequence of a certain base. Since we predict the middle base from the information on both sides, the definition of the upstream and downstream are interchangeable, which means that the context should be treated in the same way. Second, the input is quite long, far from the length $4$ of the output, which stands for the classification result of bases. The large gap between the input and output lengths leads to the fact that neural networks must be designed in a more subtle way, otherwise redundant calculations or poor results may occur. Third, we do not always have such long context data in real situations. For example, the length of upstreams in the DNA datasets in Table \ref{c3.p2.p1.tab1} mostly varies from 0 to 600, resulting in insufficient information in some cases.
\par
To solve the above problems, we designed the simple but effective input format and training method. First of all, we randomly select a certain position, taking the upstream and downstream sequence with a length of $512$ as the input, and the output is the base connecting the upstream and downstream, i.e. $\mathrm{A}$, $\mathrm{T}$, $\mathrm{C}$, $\mathrm{G}$.
\par
\begin{figure}[t]
	\centering
	\includegraphics[scale=0.92]{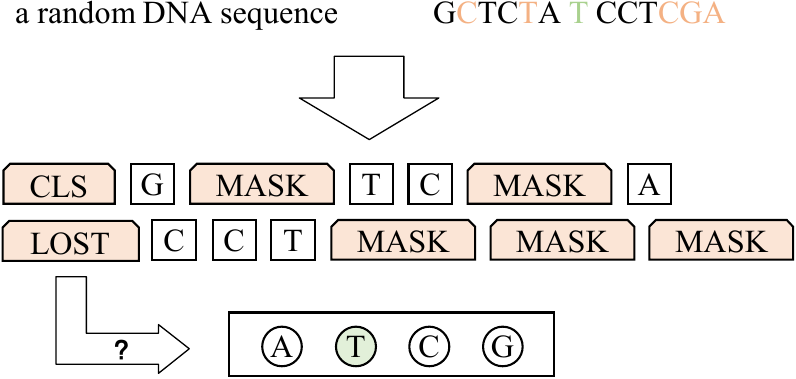}
	\caption{An example of mask operation on a random DNA sequence. Here, $\mathrm{C}$ and $\mathrm{T}$ of the upstream and $\mathrm{CGA}$ of the downstream are masked. The intermediate base is $\mathrm{T}$, which is the target needs to be predicted.}
	\label{c3.p1.p1.fig1}
\end{figure}
For the first challenge, we combine the upstream and downstream into one sequence, separated by a special prediction token $\mathrm{[LOST]}$, and provide different segment embeddings for the two parts. Also, a special classification token $\mathrm{[CLS]}$ is added to the beginning of the sequence to learn an overall representation of it.
For the second challenge, the final hidden state corresponding to the token $\mathrm{[LOST]}$ is used as the aggregate sequence representation for classification tasks, by which the output dimension is reduced quickly without complex network structures. This technique was used in Bert \citep{devlin2018bert} for the first time.
For the third challenge, some data augmentation tricks are applied in order to enhance the capabilities of the model. First, we can construct symmetric sequences based on the principle of base complementation and the non-directionality of DNA sequences, including the axial symmetry and mirror symmetry. This helps the model learn the two properties of DNA sequences. Second, we do not provide complete input to the model, which helps to enhance the model's prediction ability under insufficient information. 
Basically, we mask some percentage of the input tokens at random, and concretely there are two strategies. For a certain sequence, either upstream or downstream, we mask (replace with $\mathrm{[MASK]}$) $20\%$ of random tokens in $10\%$ of cases or $40\%$ of consecutive tokens in $15\%$ of cases. Figure \ref{c3.p1.p1.fig1} is an example below for our mask operation. In this case, $2$ random tokens of the upstream and $3$ consecutive tokens of the downstream are masked.
\par
Finally, there is no need to worry about overfitting. First, we have $10^9$ bases in the DNA dataset, meaning that we will not over-learn some specific data. Second, we have mask and dropout operations in our training, which both are great ways to avoid over-training.
\subsubsection{Model structure}
\label{c3.p1.p2}
\begin{figure}[t]
	\centering
	\includegraphics[scale=0.75]{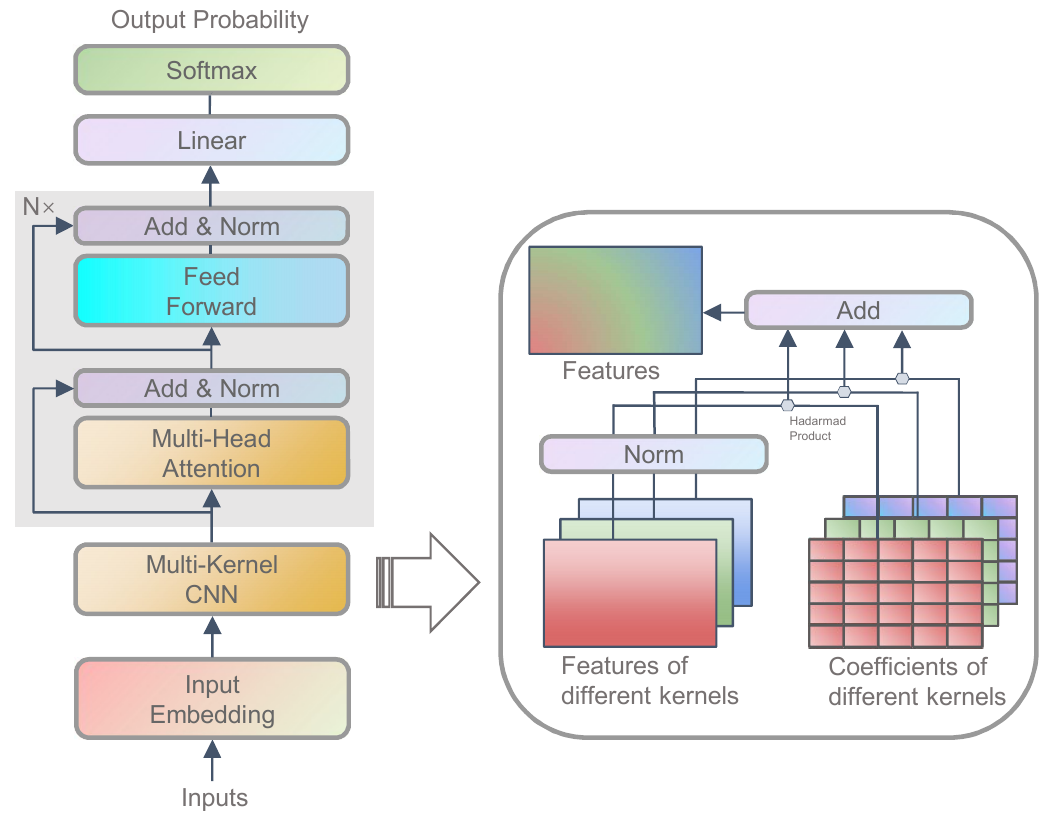}
	\caption{DNA language model. The part in the box on the right is our smooth feature addition (SFA).}
	\label{c3.p1.p2.fig1}
\end{figure}
The input and output are constructed as described in Section \ref{c3.p1.p1}, and we denote them as $T_{in}$ and $T_{out}$. Basically, the model can be described as
\begin{equation}
	\label{c3.p1.p2.eq1}
	\begin{aligned}
		T_{in}&\xrightarrow{\mathrm{embed}}T_{embed}\xrightarrow{\mathrm{multi-conv}}T_{cnns}\\
		&\xrightarrow{\mathrm{transformer}}T_{trans}\xrightarrow{\mathrm{mlp}}T_{out}
	\end{aligned}
\end{equation}
where $\mathrm{embed}$ is the Input Embedding layer, $\mathrm{multi-conv}$ stands for our Multi-Kernel CNN, $\mathrm{transformer}$ represents the transformer blocks and $\mathrm{mlp}$ contains a Linear layer and a Softmax function. Figure \ref{c3.p1.p2.fig1} shows the full picture of the model.
\par
$\mathrm{embed}$ is the encoding layer transforming the input sequence into a matrix. The dimension conversion is $\mathbb{N}^L\xrightarrow{}\mathbb{R}^{L\times E}$, Where $L$ is the length of the sequence and $E$ is the encoding length. Specifically, we encode the input as
\begin{equation}
	\label{c3.p1.p2.eq2}
	\begin{aligned}
		\mathrm{embed}(T_{in})=&\mathrm{word-embed}(T_{in})\\
		&+\mathrm{position-embed}(T_{in})\\
		&+\mathrm{segment-embed}(T_{in})
	\end{aligned}
\end{equation}
where $\mathrm{word-embed}$ is the meaning of the word itself, $\mathrm{position-embed}$ provides the representation of different positions and $\mathrm{segment-embed}$ distinguishes the upstream and the downstream. An intuitive approach is concating the three encodings without loss of semantics. But this cost triple space. Instead, we directly add these three encodings. This works because the three parameters are all leaf nodes of the training graph, which can automatically adapt to each other's distribution. By this way we reduce the dimension of the coded matrix, thus reducing the parameter space and data space.
\par
$\mathrm{multi-conv}$ is the multi-kernel convolution layer learning a short-range relationship of the sequence. The dimension conversion is $\mathbb{R}^{L\times E}\xrightarrow{}\mathbb{R}^{L\times H}$, where $H$ is the hidden dimension. Here, we use convolution kernels of different lengths to learn local relationships at different distances, and propose a smooth feature addition (SFA) method to fuse these features. Specifically, we do
\begin{equation}
	\label{c3.p1.p2.eq3}
	\mathrm{multi-conv}(T_{embed})=\sum_{i=0}^{k}\lambda_i\circ \mathrm{LN}(\mathrm{conv}_i(T_{embed}))
\end{equation}
where $\mathrm{conv}_i$ is a normal one-dimensional convolution layer with a kernel length of $l_i$, whose output dimension is $\mathbb{R}^{L\times H}$, $\lambda_i$ is a network parameter with a dimension of $\mathbb{R}^{L\times H}$ and $k$ is the number of kernels of different lengths. The sizes of convolution kernels are small rather than large ones, whose advantages have been verified in DenseNet \citep{huang2017densely}. On the one hand, small convolution kernels use less space than large convolution kernels. On the other hand, we need small convolution kernels to learn the local information of the sequence, while the long-range dependence of the sequence is to be explored by the subsequent $\mathrm{transformer}$ module.
\par
Now we explain how we designed the smooth feature addition algorithm SFA. Before that, we have to take an insight into what happens in the plain concat of features.
\par
In a sequence problem, we often directly concat two features in the last dimension. Specifically, if we have two features with dimensions $\mathbb{R}^{L\times M}$ and $\mathbb{R}^{L\times N}$, the dimension of the features after concat is $\mathbb{R}^{L\times(M+N)}$. We thought that this approach would not lose information, but in fact there is a danger of feature disappearance. For two features of different distributions learning from different modules, plain concat will bring an unbalanced distribution, where some values are extremely small. To make matters worse, layer normalization is usually used to adjust the distribution after a concat operation, making the values to be concentrated near $0$. Quantitative analysis can be seen in Theorem \ref{theorem3}. Finally, as the network goes deeper, the gradient disappears, leading to the difficulty of learning. This is proven in Corollary \ref{corollary1}.
\par
A naive thought is to normalize the two distributions before concating them, which proves to be correct in Theorem \ref{theorem1}. However, it's not effective, for it converts the dimension from $\mathbb{R}^{L\times H}$ to $\mathbb{R}^{L\times kH}$, posing a challenge for the subsequent module design. Considering that the dimensions of convolution features are the same, inspiring us to find a way to smoothly add them using some tuning parameters. That's how we designed SFA. Corollary \ref{corollary2} proves the equivalence of SFA and plain concat, and illustrates the working mechanism of SFA and its advantages in space occupation, feature selection and gradient propagation.

\par
$\mathrm{transformer}$ is the stack of transformer blocks learning a long-range relationship of the sequence. The dimension conversion is $\mathbb{R}^{L\times H}\xrightarrow{}\mathbb{R}^{L\times H}$. Simply it can be described as
\begin{equation}
	\label{c3.p1.p2.eq4}
	\mathrm{transformer}(T_{cnns})=\mathrm{sub}(\mathrm{ff}, \mathrm{sub}(\mathrm{attention}, T_{cnns}))
\end{equation}
where $\mathrm{sub}(f, x)=\mathrm{LN}(x+f(x))$. $\mathrm{ff}$ represents the feed forward function and $\mathrm{attention}$ is short for multi-head attention. The module is proposed by \citet{vaswani2017attention}.
\par
$\mathrm{mlp}$ is the output layer, responsible for converting the hidden state to the output. The dimension conversion is $\mathbb{R}^{L\times H}\xrightarrow{}\mathbb{N}$. We extract the tensor corresponding to the token $\mathrm{[LOST]}$, convert it into an output probability through a linear layer, and generate the prediction value via a softmax function. The output process is
\begin{equation}
	\label{c3.p1.p2.eq5}
	\mathrm{mlp}(T_{trans})=\mathrm{softmax}(\mathrm{linear}(T_{trans}['\mathrm{[LOST]}']))
\end{equation}
\begin{figure}[tp]
	\centering
	\includegraphics[scale=0.44]{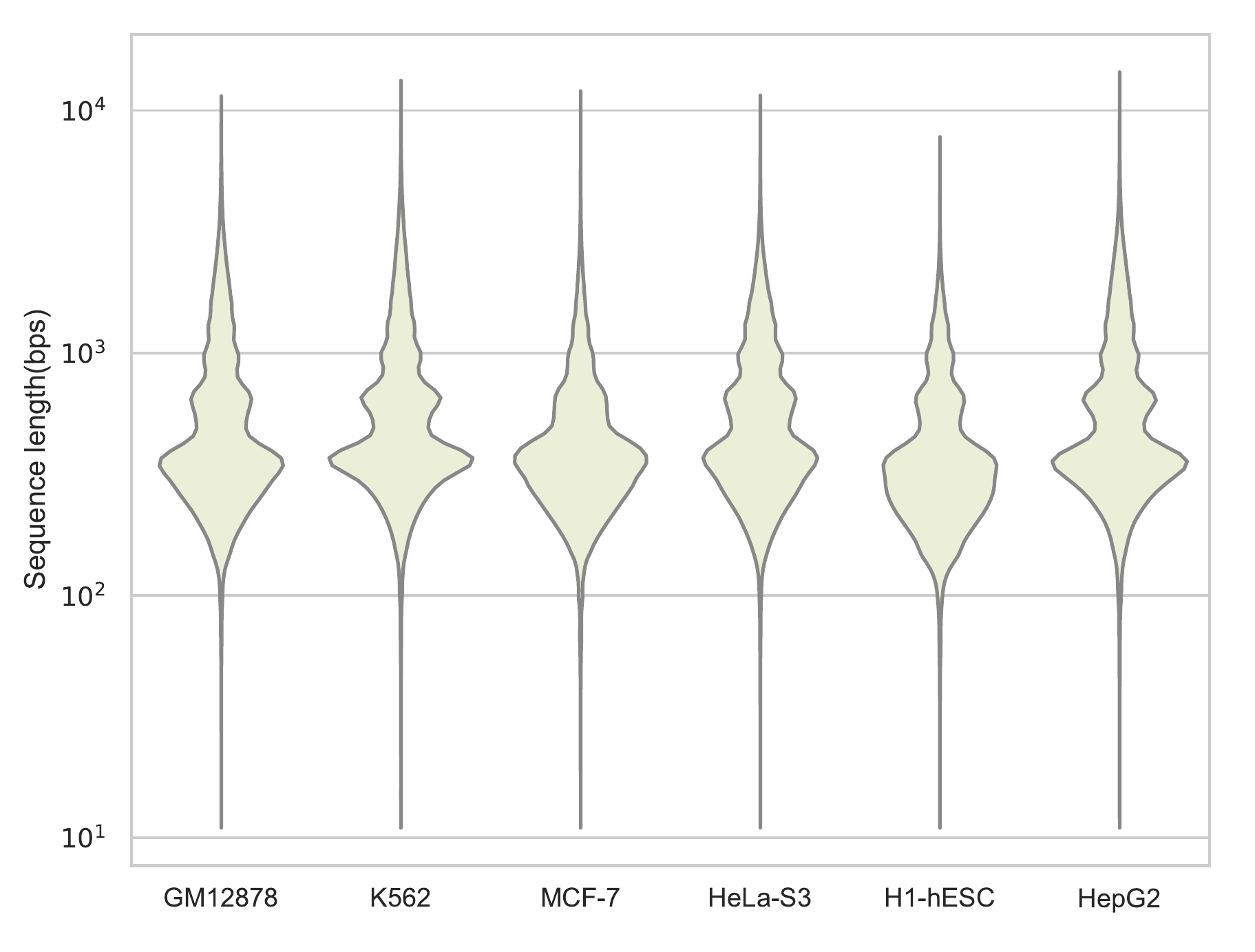}
	\caption{The length distribution of each cell, which is a more intuitive display of the content of the data. It can be seen that most of the lengths are concentrated between $10^2$-$10^3$.}
	\label{c3.p2.p1.fig2}
\end{figure}
\begin{table*}[tp]
	\centering
	\caption{An overall view of accessible DNA segments of each cell. $l$\_mean and $l$\_med show the approximate length of the data and $l$\_std describes how discrete the data are. The unit of all length is bp.}
	\setlength{\tabcolsep}{4.3mm}{
		\begin{tabular}{@{}cccccccc@{}}
			\toprule
			\textbf{Cell type} & \textbf{Code}         & \textbf{Size}   & \textbf{$l$\_min} & \textbf{$l$\_max} & \textbf{$l$\_mean} & \textbf{$l$\_med} & \textbf{$l$\_std} \\ \midrule
			GM12878   & ENCSR0000EMT & 244692 & 36     & 11481  & 610  & 381 & 614  \\
			K562      & ENCSR0000EPC & 418624 & 36     & 13307  & 675  & 423 & 671  \\
			MCF-7     & ENCSR0000EPH & 503816 & 36     & 12041  & 471  & 361 & 391  \\
			HeLa-S3   & ENCSR0000ENO & 264264 & 36     & 11557  & 615  & 420 & 524  \\
			H1-hESC   & ENCSR0000EMU & 266868 & 36     & 7795   & 430  & 320 & 347  \\
			HepG2     & ENCSR0000ENP & 283148 & 36     & 14425  & 652  & 406 & 626 \\ \bottomrule
		\end{tabular}
	}
	\label{c3.p2.p1.tab1}
\end{table*}
\subsection{Chromatin accessibility model}
\label{c3.p2}
\subsubsection{Process of data}
\label{c3.p2.p1}
We select DNase-seq experiment data of six typical cell lines as the original data for our chromatin accessibility model, including GM12878, K562, MCF-7, HeLa-S3, H1-hESC and HepG2. GM12878 is a type of lymphoblast, produced by EBV transformation from the blood of a female donor of Northern European and Western European descent. K562 is an immortalized cell derived from a female patient with chronic myeloid leukemia (CML). MCF-7 is a breast cancer cell sampled from a white female. HeLa-S3 is an immortal cell derived from a cervical cancer patient. H1-hESC is a human embryonic stem cell. HepG2 comes from a male liver cancer patient.
\par
For each cell type, we downloaded the original sequence data from the ENCODE website, use a short read aligner tool bowtie \citep{de2016bowtie} to map the DNA sequence to the human reference genome (hg19), and use HOTSPOT \citep{john2011chromatin} to identify chromatin accessibility regions (peaks). We treat these variable-length sequences as positive samples. At the same time, we sample the same number and same size sequences from the whole genome as negative samples. An overview of the data is shown in Table \ref{c3.p2.p1.tab1}, which shows the number of sequences, the minimum value, the median value, the maximum value and the standard deviation of lengths. Also, the distribution statistics of different datasets are shown in Figure \ref{c3.p2.p1.fig2}. For the fairness of comparison, we removed sequences with length less than $36$. We truncate or expand each sequence symmetrically to a sequence of length $768$, and take a context of length $512$ for each site in it. Similar to our DNA language model, a special classification token $\mathrm{[CLS]}$ is added to the beginning of the sequence to predict the accessibility. Therefore the actual input length of our model is $768+512\times2+1=1793$. From Figure \ref{c3.p2.p1.fig2}, we observe that most of the lengths are clustered between $36$ and $1792$. This proves that our cut-off has little impact and is reasonable. Compared with the input length $800$ in \cite{min2017chromatin}, our prediction length has increased by $124\%$, and the quantity of DNA sequences that do not need to be truncated in the original dataset has increased by $17.4\%$. Moreover, we do not pay a great price for such a long input because our context is handed over to a pre-trained model to predict. The output is the accessibility of the input sequence, i.e., either $0$ for inaccessibility or $1$ for accessibility.
\par
Finally, the ratio of our training set, validation set, and test set is $0.85:0.05:0.10$. The training set is used to train the model, the validation set is used to adjust the hyperparameters to prevent overfitting, and the test set is used to test the performance of the final model.
\subsubsection{Model structure}
\begin{figure}[tp]
	\centering
	\includegraphics[scale=0.70]{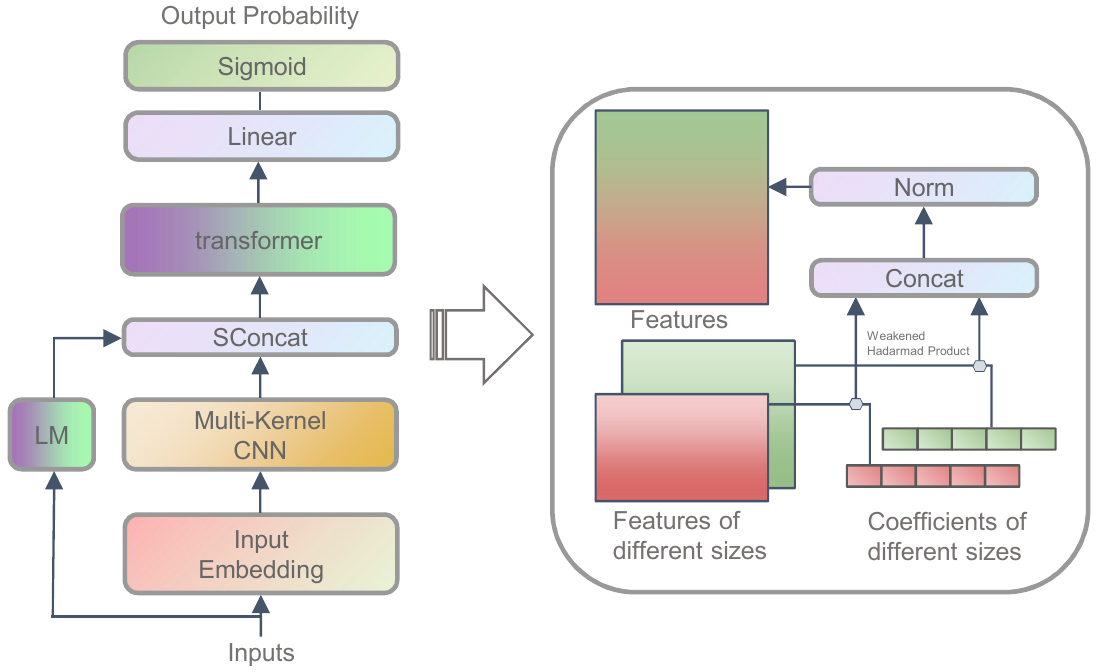}
	\caption{Chromatin accessibility model. The part in the box on the right is our smooth feature concat (SFC).}
	\label{c3.p2.p2.fig1}
\end{figure}

The input and output are constructed as described in Section \ref{c3.p2.p1}, and we denote them as $T_{in}$ and $T_{out}$. Basically, the model can be described as
\begin{equation}
	\label{c3.p2.p2.eq1}
	\begin{aligned}
	T_{in}&\xrightarrow{\mathrm{embed}}T_{embed}\xrightarrow{\mathrm{multi-conv}}T_{cnns}\xrightarrow{\mathrm{sconcat}}T_{sconcat}\\
	&\xrightarrow{\mathrm{transformer}}T_{trans}\xrightarrow{\mathrm{mlp}}T_{out}	
	\end{aligned}
\end{equation}
where $\mathrm{embed}$ is the Input Embedding layer, $\mathrm{multi-conv}$ stands for our Multi-Kernel CNN, $\mathrm{sconcat}$ is short for our SConcat module, $\mathrm{transformer}$ represents the transformer blocks and $\mathrm{mlp}$ contains a Linear layer and a Sigmoid function. Figure \ref{c3.p2.p2.fig1} shows the full picture of the model. You may find that the accessibility model is very similar to our DNA language model. Indeed, we only modify some of the model structures and change the hyperparameters, but they are all very critical adjustments which make the model suitable for the task.
\par
$\mathrm{embed}$ is the encoding layer transforming the input sequence into a feature matrix. The dimension conversion is $\mathbb{N}^L\xrightarrow{}\mathbb{R}^{L\times E}$, Where $L$ is the length of the sequence and $E$ is the encoding length. Specifically, we encode the input as
\begin{equation}
	\label{c3.p2.p2.eq2}
	\begin{aligned}
		\mathrm{embed}(T_{in})=&\mathrm{word-embed}(T_{in})\\
		&+\mathrm{position-embed}(T_{in})
	\end{aligned}
\end{equation}
Note that there is no $\mathrm{segment-embed}$ in this task because there is no need to distinguish different segments.
\par
$\mathrm{multi-conv}$ has been explained in Section \ref{c3.p1.p2}. The dimension conversion is $\mathbb{R}^{L\times E}\xrightarrow{}\mathbb{R}^{L\times G}$, where $G$ is the dimension of features learning from this layer.
\par
$\mathrm{sconcat}$ is the concat layer which fuses the features of the language model with the features learned from $\mathrm{multi-conv}$. The dimension conversion is $\mathbb{R}^{L\times G}\xrightarrow{}\mathbb{R}^{L\times (G+H)}$, where $H$ is the dimension of features generated from the DNA language model. Basically, we use the language model to construct features for sites in the sequence, and propose a smooth feature concat (SFC) method to fuse them with the previous features. What we do is
\begin{equation}
	\label{c3.p2.p2.eq3}
	\mathrm{sconcat}(T_{in}, T_{cnns})=\mathrm{LN}([\lambda_1\circ\mathrm{LM}(\overleftrightarrow{T_{in}}),\lambda_2\circ T_{cnns}])
\end{equation}
where $\overleftrightarrow{T_{in}}$ stands for the context of sites in $T_{in}$, $\lambda_1$ and $\lambda_2$ are two network parameters with a dimension of $\mathbb{R}^{L}$ and $\mathrm{LM}$ refers to our DNA language model. Here, it receives a DNA sequence, then constructs the context for each site in the sequence and produces an output of length $H$. Specifically, if the length of the sequence is $L$, it will construct $L$ pairs of context as the input and output a $\mathbb{R}^{L\times H}$ matrix.
\par
Now we explain how we designed the smooth feature concat algorithm SFC. First, we have to mention that the output dimension of the language model is $\mathbb{R}^{L\times H}$ and the dimension of $T_{cnns}$ is $\mathbb{R}^{L\times G}$, which means we cannot directly apply SFA in this scenario.
\par
Fortunately, the analysis in Section \ref{c3.p1.p2} has already given a solution. We can normalize the two distributions separately before concating them. However, this method uses $\mathrm{LN}$ twice and consumes additional parameter space and data space. One doubt is that is it possible to use $\mathrm{LN}$ only once? The answer is yes. Theorem \ref{theorem2} states that for any two distributions, there always exist two coefficients, so that the concat after they are multiplied by these two coefficients is a standardized distribution. That's how our SFA works. We multiply the two tensors by two coefficients, and then do layer normalization after the concat of them. Thereby, we fuse the two features smoothly with only one $\mathrm{LN}$ operation. Interestingly, this method is a weakened version of Theorem \ref{theorem4}.
\par
$\mathrm{transformer}$ is the same as that described in Section \ref{c3.p1.p2}. The dimension conversion is $\mathbb{R}^{L\times F}\xrightarrow{}\mathbb{R}^{L\times F}$ where $F=G+H$.
\par
$\mathrm{mlp}$ is the output layer, responsible for transforming the hidden state to the output. The dimension conversion is $\mathbb{R}^{L\times F}\xrightarrow{}\mathbb{N}$. We extract the tensor corresponding to the token $\mathrm{[CLS]}$, convert it into an output probability through a linear layer, and generate the prediction value via a sigmoid function. The output process is
\begin{equation}
	\label{c3.p2.p2.eq4}
	\mathrm{mlp}(T_{trans})=\mathrm{sigmoid}(\mathrm{linear}(T_{trans}['\mathrm{[CLS]}']))
\end{equation}

	\section{Results and Discussions}
\subsection{SemanticDNA evaluation}
\renewcommand\arraystretch{1}
\begin{table*}[tp]
	\centering
	\caption{The result of chromatin accessibility prediction system comparative experiment. The code name of these datasets can refer to Table \ref{c3.p2.p1.tab1}.}
	\setlength{\tabcolsep}{5.0mm}{
		\begin{threeparttable}
			\begin{tabular}{@{}cccccccc@{}}
				\toprule
				\textbf{System}              & \textbf{MT}         & \textbf{PC}            & \textbf{PH}           & \textbf{NO}         & \textbf{MU}         & \textbf{NP}           & \textbf{Average}         \\ \midrule
				(a) auROC           &                 &                 &                 &                 &                 &                 &                 \\
				gkmSVM              & 0.8528          & 0.8203          & 0.8967          & 0.8648          & 0.8983          & 0.8359          & 0.8697          \\
				DeepSEA             & 0.8788          & 0.8629          & 0.9200          & 0.8903          & 0.8827          & 0.8609          & 0.8782          \\
				kmer               & 0.8830          & 0.8809          & 0.9212          & 0.9016          & 0.9097          & 0.8722          & 0.8975          \\
				no feature\tnote{1}         & 0.8727          & 0.8664          & 0.9058          & 0.8840          & 0.8849          & 0.8699          & 0.8806          \\
				SemanticCAP & 0.8907 & 0.8883 & 0.9241 & 0.9001 & 0.8982 & 0.8847 & 0.8977 \\
				(b) auPRC           &                 &                 &                 &                 &                 &                 &                 \\
				gkmSVM              & 0.8442          & 0.8081          & 0.8860          & 0.8627          & 0.8823          & 0.8123          & 0.8504          \\
				DeepSEA             & 0.8758          & 0.8551          & 0.9146          & 0.8888          & 0.8705          & 0.8508          & 0.8801          \\
				kmer               & 0.8774          & 0.8732          & 0.9156          & 0.8992          & 0.8968          & 0.8630          & 0.8973          \\
				no feature\tnote         & 0.8745          & 0.8663          & 0.9053          & 0.8852          & 0.8878          & 0.8730          & 0.8820          \\
				SemanticCAP & 0.8914 & 0.8896 & 0.9218 & 0.9004 & 0.8993 & 0.8871 & 0.8983 \\ \bottomrule
			\end{tabular}
			\begin{tablenotes}
				\footnotesize
				\item[1] no feature is SemanticCAP without pre-train features.
			\end{tablenotes}
		\end{threeparttable}
	}
	\label{c4.p1.tab1}
\end{table*}
\begin{table*}[tp]
	\centering
	\caption{The result of DNA language model comparative experiment.}
	\setlength{\tabcolsep}{11.0mm}{
		\begin{threeparttable}
			\begin{tabular}{@{}cccc@{}}
				\toprule
				\textbf{Model}           & \begin{tabular}[c]{@{}c@{}}\textbf{Loss}\\\textbf{(no mask)}\end{tabular} & \begin{tabular}[c]{@{}c@{}}\textbf{Accuracy}\\\textbf{(no mask)}\end{tabular} & \begin{tabular}[c]{@{}c@{}}\textbf{Accuracy}\\\textbf{(mask $30\%$)}\end{tabular}  \\ \midrule
				convs(max)+lstms   & 1.152 & 0.4538 & 0.3265 \\
				convs(max)+attention(relu) & 1.113 & 0.4814 & 0.3687  \\
				convs(avg)+trans\tnote{1}    & 1.096 & 0.4926 & 0.3599  \\
				mconv\tnote{2} \,(PC\tnote{3} \,)+trans  & 0.968 & 0.5114 & 0.4572 \\
				mconv(PA\tnote{3} \,)+trans   & 0.931 & 0.5187 & 0.4784 \\
				mconv(SFA\tnote{3} \,)+trans   & 0.921 & 0.5202 & 0.4793 \\
				\bottomrule
			\end{tabular}
			\begin{tablenotes}
				\footnotesize
				\item[1] trans refers to transformer+linear.
				\item[2] mconv stands for our multi-conv layer.
				\item[3] PA is plain add, PC is plain concat and SFA is our smooth feature addition method.
			\end{tablenotes}
		\end{threeparttable}
	}
	\label{c4.p2.p1.tab1}
\end{table*}
\begin{table*}[tp]
	\centering
	\caption{The result of chromatin accessibility model comparative experiment.}
	\setlength{\tabcolsep}{5.3mm}{
		\begin{threeparttable}
			\begin{tabular}{@{}cccccc@{}}
				\toprule
				\textbf{Model} & \begin{tabular}[c]{@{}c@{}}\textbf{Parameters}\\ \textbf{(M)}\end{tabular}  & \begin{tabular}[c]{@{}c@{}}\textbf{Total}\\ \textbf{(h)}\end{tabular} & \textbf{auROC}    & \textbf{auPRC}  & \textbf{\quad F1\quad}      \\ \midrule
				PC+trans\tnote{1} +lstm         & 4.16        & 4.6     & 0.8595 & 0.8625 & 0.7880 \\
				PC+trans+conv+lstm    & 4.95          & 2.9      & 0.8741 & 0.8765 & 0.8036 \\
				PC+trans+flatten      & 16.4        & 2.0      & 0.8822 & 0.8839 & 0.8124 \\
				PC+trans+conv+flatten & 6.13           & 2.8       & 0.8817 & 0.8834 & 0.8119 \\
				PC+trans+linear  & 3.84 & 1.5 & 0.8839 & 0.8854 & 0.8144 \\
				mconv+PC+trans+linear        & 5.61       & 2.5     & 0.8881 & 0.8902 & 0.8590 \\ 
				mconv+SFC\tnote{2} +trans +linear            &  5.61      & 2.5     & 0.8907 & 0.8914 & 0.8606 \\ 
				\bottomrule
			\end{tabular}
			\begin{tablenotes}
				\footnotesize
				\item[1] trans is short for transformer blocks.
				\item[2] SFC is our smooth feature concat method.
			\end{tablenotes}
		\end{threeparttable}
	}
	\label{c4.p2.p2.tab1}
\end{table*}
\renewcommand\arraystretch{1}
We compared the performance of our proposed method with several baseline methods, including the gkmSVM \citep{ghandi2014enhanced}, DeepSEA \citep{zhou2015predicting} and kmer \citep{min2017chromatin}. For the sake of fairness, all parameters were set as default. Besides, to prove the effectiveness of the DNA language model, we also tested our accessibility model excluding the DNA language model. For evaluation purpose, we computed two often-used measures, the area under the receiver operating characteristic curve (auROC) and the area under the precision-recall curve (auPRC), which are good indicators of the robustness of a prediction model. The classification results on six datasets are shown in Table \ref{c4.p1.tab1}. Compared to the best baseline kmer, our system has a maximum $1.25\%$ improvement in auROC, and a maximum $2.41\%$ improvement in auPRC. Although some results on some datasets are not good, our model outperforms kmer on average, with $0.02\%$ higher auROC score and $0.1\%$ higher auPRC score. Compared to gkmSVM and DeepSEA, SemanticCAP has an about $2\%$-$3\%$ improvement on average. Finally, the introduction of our DNA language model brings about a $2\%$ performance improvement.
\par
We also tested the accessibility prediction accuracy of loci shared in different cell lines. For example, GM12878 and HeLa-S3 have $20$ common loci, and the prediction accuracy of these 20 loci in both cell lines is $85\%$ and $90\%$. Another example is that K562 and MCF-7 have $21$ common loci, and the prediction accuracy is $80.9\%$ and $90.5\%$, respectively. This shows the applicability of our system on the common loci between different cell lines.
\subsection{Analysis of models}
\subsubsection{Effectiveness of our DNA language model}
We did experiments on several different DNA language model structures, which can be roughly divided into two categories.
The first category can be attributed to methods based on normal CNNs and the second uses our multi-conv architecture with data augmentation. Six structures are tested. At the same time, in order to test the prediction ability of different models in the case of insufficient information, we randomly masked some words and test the results. Complete results are shown in Table \ref{c4.p2.p1.tab1}. Through the comparison of LSTM and Attention, we can find that the attention mechanism can greatly improve the prediction ability of the DNA language model. When using the MaxPooling and ReLU functions, we observed that the output of the last hidden layer is mostly $0$, where the number of effective (not zero) neurons is about $3/256$. This is because the ReLU function shields neurons whose values are less than $0$, and MaxPooling selectively updates specific neurons. Therefore, we replace the MaxPooling with the AveragePooling, and the Attention layer that uses the ReLU function is replaced with a transformer. That's the third method in Table \ref{c4.p2.p1.tab1}.
The second category uses multi-conv to extract local features of the sequence. The introduction to multi-conv mechanism with data augmentation strategies brings increasement in accuracy, especially when we mask some tokens. There are three kinds of feature fusion strategies: plain concat (PC), plain add (PA) and our smooth feature add (SFA). The third, fourth, and fifth items in the table indicate that SFA outperforms the other two fusion methods. The last item in Table \ref{c4.p2.p1.tab1}, mconv(SFA)+trans, is the model we finally chose as our DNA language model. 
\par
\subsubsection{Effectiveness of our chromatin accessibility model}
We experimented with several chromatin accessibility model structures, which are all based on the transformer. The main difference is the use of multi-conv and the modules after the transformer. Complete comparison results are shown in Table \ref{c4.p2.p2.tab1}.
\par
First focus on the module before the transformer. We notice that the introduction of multi-conv also brought performance improvements, especially in F1. In our chromatin accessibility model, we concat the features provided by the DNA language model, where we can either directly concat (PC) them or using our SFC method. The evaluation values of the last two items show the superiority of SFC.
\par
Now turn to the comparison of modules after the transformer. The transformation from the features of transformer to the result is a challenge. In this part, five methods are tested. Mention that mconv+SFC+trans+linear is the final model. In terms of training time, our model can be fully parallelized, which is more advantageous than LSTM based on recurrent networks. At the same time, our model has fewer parameters and a simpler structure than Flatten after CNNs, thus can converge quickly. In terms of evaluation, LSTM-based methods perform poorly, and the main reason is that it is difficult for LSTM to learn the long-range dependence of a sequence. The convolution layer improves the performance of the LSTM to some extent by shortening the sequence length. In methods based on Flatten, the introduction of convolution layers actually reduces the accuracy. Maybe it's because that the convolution layers destroy the sequence features learned from the transformer.
During multiple chromatin accessibility models, the method using mult-conv and our smoother concat method SFC obtain the best results with relatively small number of parameters.
\subsection{Analysis of $\mathrm{[CLS]}$}
\begin{figure}[tp]
	\centering
	\includegraphics[scale=0.44]{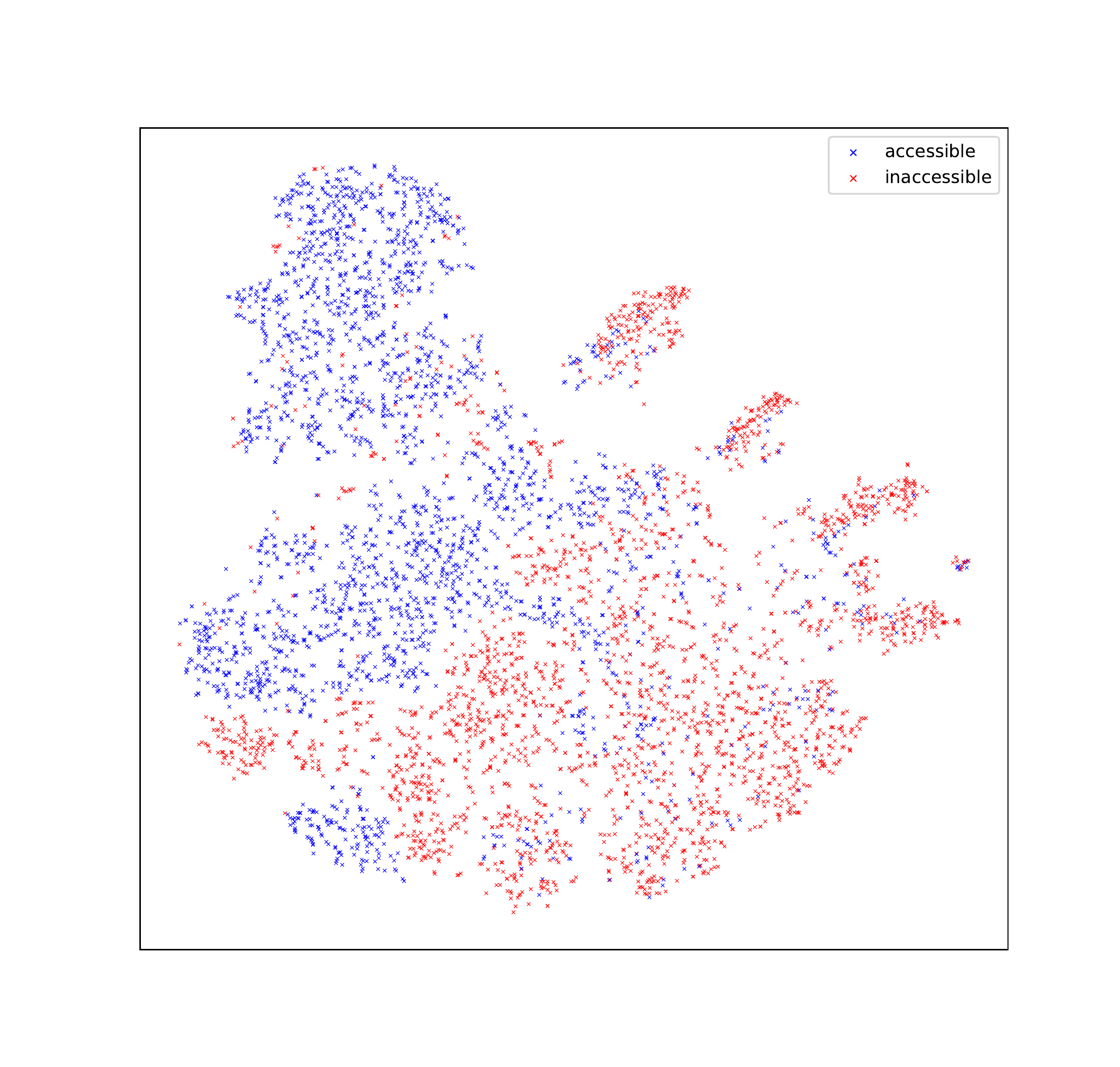}
	\caption{Features corresponding to the token $\mathrm{[CLS]}$ after the transformer for different samples. Accessible points and inaccessible points can be roughly distinguished.}
	\label{c4.p3.fig1}
\end{figure}
We can observe the effectiveness of introducing the $\mathrm{[CLS]}$ symbol in our accessibility model. A direct indicator is the feature corresponding to $\mathrm{[CLS]}$ after the transformer layer, i.e., the value of $T_{trans}['\mathrm{[CLS]}']$ in Eq. \ref{c3.p2.p2.eq4}. We randomly selected a certain number of positive and negative samples and use our chromatin accessibility model to predict them. For each sample, we output the $256$-dimensional tensor corresponding to $\mathrm{[CLS]}$ after the transformer layer, and reduced it to two-dimensional space with t-SNE, which is shown in Figure \ref{c4.p3.fig1}. According to it, the feature has an ability to distinguish positive and negative examples, which is strong evidence of its effectiveness.
\subsection{Analysis of $\mathrm{SFA}$ and $\mathrm{SFC}$}
In this part we did two comparison experiments of PA, PC, SFA and SFC.
\par
\begin{figure}[tp]
	\centering
	\includegraphics[scale=0.30]{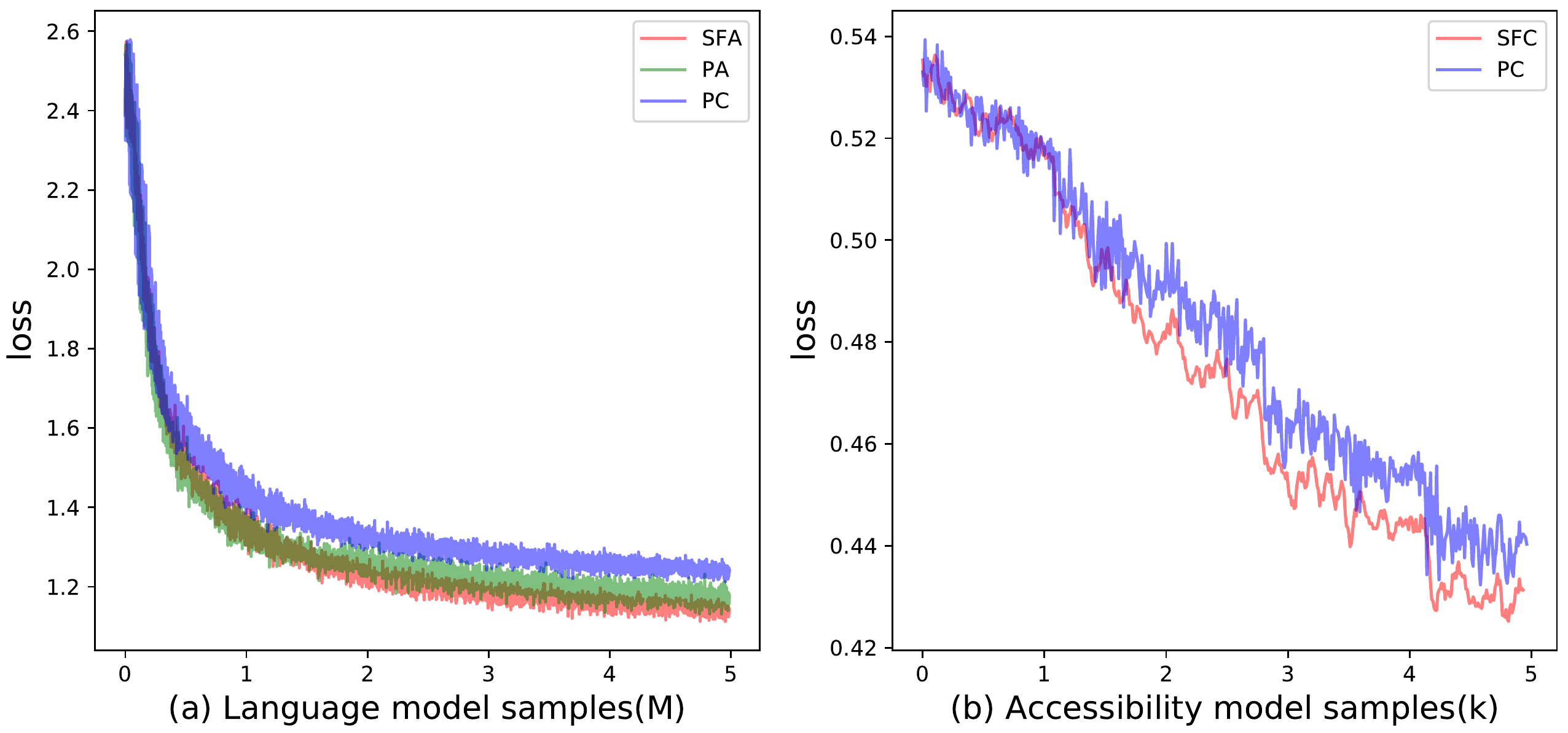}
	\caption{Loss in training time. \textbf{(a)} The loss curve for SFA, PA and PC in the training of DNA language model. \textbf{(b)} The loss curve for SFC and PC in the training of chromatin accessibility model.}
	\label{c4.p4.fig1}
\end{figure}

In the experiment of various DNA language models, we made a comparison between SFA, PA and PC, corresponding to the last three items in Table \ref{c4.p2.p1.tab1}. We used $5\times 10^6$ samples to train the three models, drew the training loss map of them and saw what would happen, which is shown in Figure \ref{c4.p4.fig1}a. PA quickly reduces the loss at the fastest speed at the beginning, because all features in multi-conv are trained at the same time to the same degree. But in the later stage, there appears a phenomenon that some features are overtrained, while others are not, leading to the oscillation of loss.
\par
In the experiment of various chromatin accessibility models, we made a comparison between SFC and PC, corresponding to the last two items in Table \ref{c4.p2.p2.tab1}. The first $5\times 10^3$ samples are used to measure its training state, which is shown in Figure \ref{c4.p4.fig1}b. According to that, PC has a lower training speed, for it has a problem of gradient disappearance. Compared to it, the gradient propagation of SFA is selective and more stable for the whole term.
\par
\begin{figure}[tp]
	\centering
	\includegraphics[scale=0.30]{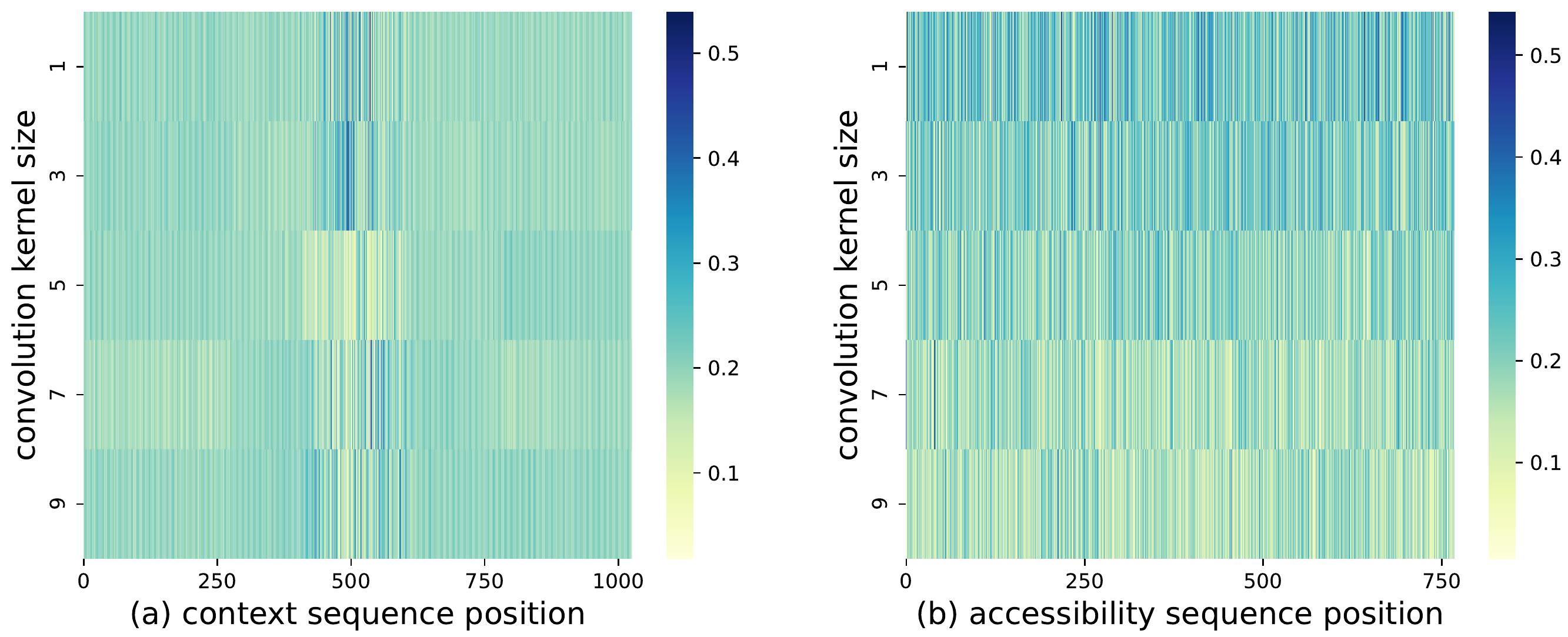}
	\caption{SFA and SFC parameters. \textbf{(a)} Parameters of SFA in the DNA language model. \textbf{(b)} Parameters of SFC in the chromatin accessibility model. Different sequence positions and different convolution kernels have different weights, which proves SFA’s ability to regulate features.}
	\label{c4.p4.fig2}
\end{figure}

We can observe the effectiveness of SFA from another angle. Pay attention to the parameters $C_{SFA}$ of SFA in multi-conv, whose dimension is $\mathbb{R}^{K\times L\times H}$, where $K$ is the number of kernels, $L$ is the sequence length and $H$ is the hidden dimension. We normalized it and converted it to $C'_{SFA}$, whose dimension is $\mathbb{R}^{K\times L}$. We do this for both the language model and the chromatin accessibility model, and picture them on Figure \ref{c4.p4.fig2}. Note that the sum of the vertical axis is always $1$ due to the normalization. Obviously, different sequence positions and different convolution kernels have different weights, which proves SFA's ability to regulate features.
\par
In general, SFA and SFC make training smoother, faster, and better. They are simple but effective. Actually, since they share the same essence (Hadamard product), they share the same advantages.

	\section{Conclusion}
In this article, we propose a chromatin accessibility prediction model named SemanticCAP. Our model is able to predict open regions of DNA, thus having a guiding role in disease detection, drug design, etc. For example, a gene called \textit{CYMC} from cell H1-hESC mutated in the middle with a length of 5 bp, and its accessibility decreased from $0.98$ to $0.14$ predicted by our model, which is consistent with the experimental data that it reduces transcription \citep{klenova1993ctcf}. Another example is a mutation in a gene called \textit{HNF4A} from cell K562, which leads to a reduction in gene expression \citep{ellard2006mutations}. Our model predicted that its accessibility decreased from $0.66$ to $0.2$, which gives a reasonable explanation for the experimental phenomenon of reduction in gene expression caused by the mutation. Similarly, we can monitor the accessibility changes of DNA targeted by drugs (especially anticancer drugs), and the change of accessibility will provide guidance for drug action. The innovations of our method are as follows. First, we introduce the concept of language models in natural language processing to model DNA sequences. This method not only provides the word vector presentation of the base itself, but also provides sufficient information about the context of a site in a DNA sequence. Second, we use a small number of parameters to solve the feature fusion problem between different distributions. Specifically, we solve the problem of smooth addition of same dimensional distributions using SFA and the problem of smooth concat of different dimensional distributions using SFC. Third, we use an end-to-end model design, in which we fully utilize the learning ability and characteristics of the convolution and attention mechanism, thus achieving a better result with fewer parameters and shorter training time.
\par
Of course, there is still much room for improvement in our method. In terms of the sample construction, we randomly select the same number of DNA sequences with the same length as negative samples. This approach may be modified. For example, we can deliberately use an unbalanced dataset, for there are so much DNA data, and then use some strategies such as ensemble learning \citep{dietterich2002ensemble} to eliminate the negative effects of data imbalance \citep{chawla2007exploiting}. In terms of data input, sequence trucation or sequence completion operations exist in our model, which may cause information loss or redundant calculation. Also, the task we designed for the DNA language model could be enhanced. Multiple positions can be predicted simultaneously, just like the cloze problem in Bert. There are many other potential improvements. First is that the attention mechanism consumes too much memory, which could be replaced by a short-range attention or a mixed-length attention \citep{choromanski2020rethinking}. Also, our smooth feature fusion method SFA and SFC could also be used in the multi-head attention to save space and accelerate training. Additionally, the dropout mechanism makes all neurons effective in the prediction phase, but there may exist a more reasonable way of fusing sub-networks. These issues need to be further explored.
	\section*{Acknowledgement}
This work was supported by the following: the National Natural Science Foundation of China (31801108, 62002251); the Natural Science Foundation of Jiangsu Province Youth Fund (BK20200856); and A Project Funded by the Priority Academic Program Development of Jiangsu Higher Education Institutions (PAPD). This work was partially supported by the Collaborative Innovation Center of Novel Software Technology and Industrialization. The authors would also like to thank the support of Jiangsu Province Key Lab for providing information processing technologies.
\bibliography{ref.bib}	

\begin{thebibliography}{38}
\providecommand{\natexlab}[1]{#1}
\expandafter\ifx\csname urlstyle\endcsname\relax
  \providecommand{\doi}[1]{doi:\discretionary{}{}{}#1}\else
  \providecommand{\doi}{doi:\discretionary{}{}{}\begingroup
  \urlstyle{rm}\Url}\fi
\providecommand{\selectlanguage}[1]{\relax}
\providecommand{\bibAnnoteFile}[1]{%
  \IfFileExists{#1}{\begin{quotation}\noindent\textsc{Key:} #1\\
  \textsc{Annotation:}\ \input{#1}\end{quotation}}{}}
\providecommand{\bibAnnote}[2]{%
  \begin{quotation}\noindent\textsc{Key:} #1\\
  \textsc{Annotation:}\ #2\end{quotation}}

\bibitem[{Aleksi{\'c} and Kapetanovi{\'c}(2014)}]{aleksic2014overview}
Aleksi{\'c}, M. and Kapetanovi{\'c}, V. (2014).
\newblock An overview of the optical and electrochemical methods for detection
  of dna-drug interactions.
\newblock \emph{Acta Chimica Slovenica} 61, 555--573
\bibAnnoteFile{aleksic2014overview}

\bibitem[{Alipanahi et~al.(2015)Alipanahi, Delong, Weirauch, and
  Frey}]{alipanahi2015predicting}
Alipanahi, B., Delong, A., Weirauch, M.~T., and Frey, B.~J. (2015).
\newblock Predicting the sequence specificities of dna-and rna-binding proteins
  by deep learning.
\newblock \emph{Nature biotechnology} 33, 831--838
\bibAnnoteFile{alipanahi2015predicting}

\bibitem[{Ba et~al.(2016)Ba, Kiros, and Hinton}]{ba2016layer}
Ba, J.~L., Kiros, J.~R., and Hinton, G.~E. (2016).
\newblock Layer normalization.
\newblock \emph{arXiv preprint arXiv:1607.06450}
\bibAnnoteFile{ba2016layer}

\bibitem[{Baldi and Sadowski(2013)}]{baldi2013understanding}
Baldi, P. and Sadowski, P.~J. (2013).
\newblock Understanding dropout.
\newblock \emph{Advances in neural information processing systems} 26,
  2814--2822
\bibAnnoteFile{baldi2013understanding}

\bibitem[{Beer(2017)}]{beer2017predicting}
Beer, M.~A. (2017).
\newblock Predicting enhancer activity and variant impact using gkm-svm.
\newblock \emph{Human mutation} 38, 1251--1258
\bibAnnoteFile{beer2017predicting}

\bibitem[{Bengio et~al.(2013)Bengio, Courville, and
  Vincent}]{bengio2013representation}
Bengio, Y., Courville, A., and Vincent, P. (2013).
\newblock Representation learning: A review and new perspectives.
\newblock \emph{IEEE transactions on pattern analysis and machine intelligence}
  35, 1798--1828
\bibAnnoteFile{bengio2013representation}

\bibitem[{Buenrostro et~al.(2015)Buenrostro, Wu, Chang, and
  Greenleaf}]{buenrostro2015atac}
Buenrostro, J.~D., Wu, B., Chang, H.~Y., and Greenleaf, W.~J. (2015).
\newblock Atac-seq: a method for assaying chromatin accessibility genome-wide.
\newblock \emph{Current protocols in molecular biology} 109, 21--29
\bibAnnoteFile{buenrostro2015atac}

\bibitem[{Chawla and Sylvester(2007)}]{chawla2007exploiting}
Chawla, N.~V. and Sylvester, J. (2007).
\newblock Exploiting diversity in ensembles: Improving the performance on
  unbalanced datasets.
\newblock In \emph{International Workshop on Multiple Classifier Systems}
  (Springer), 397--406
\bibAnnoteFile{chawla2007exploiting}

\bibitem[{Choromanski et~al.(2020)Choromanski, Likhosherstov, Dohan, Song,
  Gane, Sarlos et~al.}]{choromanski2020rethinking}
Choromanski, K., Likhosherstov, V., Dohan, D., Song, X., Gane, A., Sarlos, T.,
  et~al. (2020).
\newblock Rethinking attention with performers.
\newblock \emph{arXiv preprint arXiv:2009.14794}
\bibAnnoteFile{choromanski2020rethinking}

\bibitem[{de~Ruijter and Guldenmund(2016)}]{de2016bowtie}
de~Ruijter, A. and Guldenmund, F. (2016).
\newblock The bowtie method: A review.
\newblock \emph{Safety science} 88, 211--218
\bibAnnoteFile{de2016bowtie}

\bibitem[{Devlin et~al.(2018)Devlin, Chang, Lee, and
  Toutanova}]{devlin2018bert}
Devlin, J., Chang, M.-W., Lee, K., and Toutanova, K. (2018).
\newblock Bert: Pre-training of deep bidirectional transformers for language
  understanding.
\newblock \emph{arXiv preprint arXiv:1810.04805}
\bibAnnoteFile{devlin2018bert}

\bibitem[{Dietterich et~al.(2002)}]{dietterich2002ensemble}
Dietterich, T.~G. et~al. (2002).
\newblock Ensemble learning.
\newblock \emph{The handbook of brain theory and neural networks} 2, 110--125
\bibAnnoteFile{dietterich2002ensemble}

\bibitem[{Ellard and Colclough(2006)}]{ellard2006mutations}
Ellard, S. and Colclough, K. (2006).
\newblock Mutations in the genes encoding the transcription factors hepatocyte
  nuclear factor 1 alpha (hnf1a) and 4 alpha (hnf4a) in maturity-onset diabetes
  of the young.
\newblock \emph{Human mutation} 27, 854--869
\bibAnnoteFile{ellard2006mutations}

\bibitem[{Gallon et~al.(2021)Gallon, Loomis, Curry, Martin, Brody, Garner
  et~al.}]{gallon2021chromatin}
Gallon, J., Loomis, E., Curry, E., Martin, N., Brody, L., Garner, I., et~al.
  (2021).
\newblock Chromatin accessibility changes at intergenic regions are associated
  with ovarian cancer drug resistance.
\newblock \emph{Clinical epigenetics} 13, 1--15
\bibAnnoteFile{gallon2021chromatin}

\bibitem[{Ghandi et~al.(2014)Ghandi, Lee, Mohammad-Noori, and
  Beer}]{ghandi2014enhanced}
Ghandi, M., Lee, D., Mohammad-Noori, M., and Beer, M.~A. (2014).
\newblock Enhanced regulatory sequence prediction using gapped k-mer features.
\newblock \emph{PLoS computational biology} 10, e1003711
\bibAnnoteFile{ghandi2014enhanced}

\bibitem[{Gin{\'e}(1983)}]{gine1983levy}
Gin{\'e}, E. (1983).
\newblock The l{\'e}vy-lindeberg central limit theorem.
\newblock \emph{Proceedings of the American Mathematical Society} 88, 147--153
\bibAnnoteFile{gine1983levy}

\bibitem[{Guo et~al.(2020)Guo, Zhou, Nie, Ruan, and Li}]{guo2020deepanf}
Guo, Y., Zhou, D., Nie, R., Ruan, X., and Li, W. (2020).
\newblock Deepanf: A deep attentive neural framework with distributed
  representation for chromatin accessibility prediction.
\newblock \emph{Neurocomputing} 379, 305--318
\bibAnnoteFile{guo2020deepanf}

\bibitem[{Hochreiter and Schmidhuber(1997)}]{hochreiter1997long}
Hochreiter, S. and Schmidhuber, J. (1997).
\newblock Long short-term memory.
\newblock \emph{Neural computation} 9, 1735--1780
\bibAnnoteFile{hochreiter1997long}

\bibitem[{Horn(1990)}]{horn1990hadamard}
Horn, R.~A. (1990).
\newblock The hadamard product.
\newblock In \emph{Proc. Symp. Appl. Math}. vol.~40, 87--169
\bibAnnoteFile{horn1990hadamard}

\bibitem[{Huang et~al.(2017)Huang, Liu, Van Der~Maaten, and
  Weinberger}]{huang2017densely}
Huang, G., Liu, Z., Van Der~Maaten, L., and Weinberger, K.~Q. (2017).
\newblock Densely connected convolutional networks.
\newblock In \emph{Proceedings of the IEEE conference on computer vision and
  pattern recognition}. 4700--4708
\bibAnnoteFile{huang2017densely}

\bibitem[{Janssen et~al.(2000)Janssen, Cuvier, M{\"u}ller, and
  Laemmli}]{janssen2000specific}
Janssen, S., Cuvier, O., M{\"u}ller, M., and Laemmli, U.~K. (2000).
\newblock Specific gain-and loss-of-function phenotypes induced by
  satellite-specific dna-binding drugs fed to drosophila melanogaster.
\newblock \emph{Molecular cell} 6, 1013--1024
\bibAnnoteFile{janssen2000specific}

\bibitem[{John et~al.(2011)John, Sabo, Thurman, Sung, Biddie, Johnson
  et~al.}]{john2011chromatin}
John, S., Sabo, P.~J., Thurman, R.~E., Sung, M.-H., Biddie, S.~C., Johnson,
  T.~A., et~al. (2011).
\newblock Chromatin accessibility pre-determines glucocorticoid receptor
  binding patterns.
\newblock \emph{Nature genetics} 43, 264--268
\bibAnnoteFile{john2011chromatin}

\bibitem[{Kalchbrenner et~al.(2014)Kalchbrenner, Grefenstette, and
  Blunsom}]{kalchbrenner2014convolutional}
Kalchbrenner, N., Grefenstette, E., and Blunsom, P. (2014).
\newblock A convolutional neural network for modelling sentences.
\newblock \emph{arXiv preprint arXiv:1404.2188}
\bibAnnoteFile{kalchbrenner2014convolutional}

\bibitem[{Klenova et~al.(1993)Klenova, Nicolas, Paterson, Carne, Heath, Goodwin
  et~al.}]{klenova1993ctcf}
Klenova, E., Nicolas, R., Paterson, H., Carne, A., Heath, C., Goodwin, G.,
  et~al. (1993).
\newblock Ctcf, a conserved nuclear factor required for optimal transcriptional
  activity of the chicken c-myc gene, is an 11-zn-finger protein differentially
  expressed in multiple forms.
\newblock \emph{Molecular and cellular biology} 13, 7612--7624
\bibAnnoteFile{klenova1993ctcf}

\bibitem[{Kumar and Bucher(2016)}]{kumar2016predicting}
Kumar, S. and Bucher, P. (2016).
\newblock Predicting transcription factor site occupancy using dna sequence
  intrinsic and cell-type specific chromatin features.
\newblock In \emph{BMC bioinformatics} (Springer), vol.~17, 41--50
\bibAnnoteFile{kumar2016predicting}

\bibitem[{Lee et~al.(2011)Lee, Karchin, and Beer}]{lee2011discriminative}
Lee, D., Karchin, R., and Beer, M.~A. (2011).
\newblock Discriminative prediction of mammalian enhancers from dna sequence.
\newblock \emph{Genome research} 21, 2167--2180
\bibAnnoteFile{lee2011discriminative}

\bibitem[{Liu and Perez(2017)}]{liu2017gated}
Liu, F. and Perez, J. (2017).
\newblock Gated end-to-end memory networks.
\newblock In \emph{Proceedings of the 15th Conference of the European Chapter
  of the Association for Computational Linguistics: Volume 1, Long Papers}.
  1--10
\bibAnnoteFile{liu2017gated}

\bibitem[{Liu et~al.(2018)Liu, Xia, Yin, and Jiang}]{liu2018chromatin}
Liu, Q., Xia, F., Yin, Q., and Jiang, R. (2018).
\newblock Chromatin accessibility prediction via a hybrid deep convolutional
  neural network.
\newblock \emph{Bioinformatics} 34, 732--738
\bibAnnoteFile{liu2018chromatin}

\bibitem[{Min et~al.(2017)Min, Zeng, Chen, Chen, and Jiang}]{min2017chromatin}
Min, X., Zeng, W., Chen, N., Chen, T., and Jiang, R. (2017).
\newblock Chromatin accessibility prediction via convolutional long short-term
  memory networks with k-mer embedding.
\newblock \emph{Bioinformatics} 33, i92--i101
\bibAnnoteFile{min2017chromatin}

\bibitem[{Pan et~al.(2019)Pan, Kusko, Xiao, Zheng, Liu, Xiao
  et~al.}]{pan2019similarities}
Pan, B., Kusko, R., Xiao, W., Zheng, Y., Liu, Z., Xiao, C., et~al. (2019).
\newblock Similarities and differences between variants called with human
  reference genome hg19 or hg38.
\newblock \emph{BMC bioinformatics} 20, 17--29
\bibAnnoteFile{pan2019similarities}

\bibitem[{Pennington et~al.(2014)Pennington, Socher, and
  Manning}]{pennington2014glove}
Pennington, J., Socher, R., and Manning, C.~D. (2014).
\newblock Glove: Global vectors for word representation.
\newblock In \emph{Proceedings of the 2014 conference on empirical methods in
  natural language processing (EMNLP)}. 1532--1543
\bibAnnoteFile{pennington2014glove}

\bibitem[{Simon et~al.(2012)Simon, Giresi, Davis, and Lieb}]{simon2012using}
Simon, J.~M., Giresi, P.~G., Davis, I.~J., and Lieb, J.~D. (2012).
\newblock Using formaldehyde-assisted isolation of regulatory elements (faire)
  to isolate active regulatory dna.
\newblock \emph{Nature protocols} 7, 256--267
\bibAnnoteFile{simon2012using}

\bibitem[{Song and Crawford(2010)}]{song2010dnase}
Song, L. and Crawford, G.~E. (2010).
\newblock Dnase-seq: a high-resolution technique for mapping active gene
  regulatory elements across the genome from mammalian cells.
\newblock \emph{Cold Spring Harbor Protocols} 2010, pdb--prot5384
\bibAnnoteFile{song2010dnase}

\bibitem[{Sun et~al.(2019)Sun, Yang, Luo, Wang, Zhang, Lin
  et~al.}]{sun2019deep}
Sun, C., Yang, Z., Luo, L., Wang, L., Zhang, Y., Lin, H., et~al. (2019).
\newblock A deep learning approach with deep contextualized word
  representations for chemical--protein interaction extraction from biomedical
  literature.
\newblock \emph{IEEE Access} 7, 151034--151046
\bibAnnoteFile{sun2019deep}

\bibitem[{Vaswani et~al.(2017)Vaswani, Shazeer, Parmar, Uszkoreit, Jones, Gomez
  et~al.}]{vaswani2017attention}
Vaswani, A., Shazeer, N., Parmar, N., Uszkoreit, J., Jones, L., Gomez, A.~N.,
  et~al. (2017).
\newblock Attention is all you need.
\newblock In \emph{Advances in neural information processing systems}.
  5998--6008
\bibAnnoteFile{vaswani2017attention}

\bibitem[{Wang et~al.(2016)Wang, Jiang, and Wong}]{wang2016modeling}
Wang, Y., Jiang, R., and Wong, W.~H. (2016).
\newblock Modeling the causal regulatory network by integrating chromatin
  accessibility and transcriptome data.
\newblock \emph{National science review} 3, 240--251
\bibAnnoteFile{wang2016modeling}

\bibitem[{Xu and Strick(2021)}]{xu2021integration}
Xu, Y. and Strick, A.~J. (2021).
\newblock Integration of unpaired single-cell chromatin accessibility and gene
  expression data via adversarial learning.
\newblock \emph{arXiv preprint arXiv:2104.12320}
\bibAnnoteFile{xu2021integration}

\bibitem[{Zhou and Troyanskaya(2015)}]{zhou2015predicting}
Zhou, J. and Troyanskaya, O.~G. (2015).
\newblock Predicting effects of noncoding variants with deep learning--based
  sequence model.
\newblock \emph{Nature methods} 12, 931--934
\bibAnnoteFile{zhou2015predicting}

\end{thebibliography}
\end{document}